\documentclass[amsfonts,preprintnumbers,showpacs,twocolumn]{revtex4-1}
\usepackage[applemac]{inputenc}
\usepackage{graphicx,hyperref,mathptmx,svn}
\pdfoutput=1

\newcommand{\CCC}{Cs$_2$CuCl$_4$}
\newcommand{\CCB}{Cs$_2$CuBr$_4$}
\newcommand{\CCCB}{Cs$_2$CuCl$_{4-x}$Br$_x$}

\def\vec#1{{\bf #1}}

\SVN $Id: lsw-triangle.tex 50 2014-04-12 14:46:58Z bs $

\begin{document}

\title{Quantum fluctuations in anisotropic triangular lattices with
ferro- and antiferromagnetic exchange}

\author{Burkhard Schmidt and Peter Thalmeier}
\affiliation{Max-Planck-Institut f{\"u}r Chemische Physik fester
Stoffe, 01187 Dresden, Germany}

\begin{abstract}
    The Heisenberg model on a triangular lattice is a prime example
    for a geometrically frustrated spin system.  However most 
    experimentally accessible compounds have spatially anisotropic
    exchange interactions.  As a function of this anisotropy, ground
    states with different magnetic properties can be realized.
    Motivated by recent experimental findings on
    Cs$_{2}$CuCl$_{4-x}$Br$_{x}$, we discuss the full phase diagram of
    the anisotropic model with two exchange constants $J_{1}$ and
    $J_{2}$, including possible ferromagnetic exchange. Furthermore a
    comparison with the related square lattice model is carried out.
We discuss
    the zero-temperature phase diagram, ordering vector, ground-state
    energy, and ordered moment on a classical level and investigate
the
    effect of quantum fluctuations within the framework of spin-wave
    theory. The field dependence of the ordered moment is shown to be
    nonmonotonic with field and control parameter.
\end{abstract}


%\preprint{\SVNId}

\maketitle

\section{Introduction}
\label{sec:introduction}

The 2D isotropic triangular Heisenberg antiferromagnet (HAF) is the
most simple spin model that exhibits geometric frustration when only
next neighbor (n.n) coupling is considered.  In fact the triangular
lattice is the only Bravais lattice in 2D which is geometrically
frustrated.  Classically (i.e., for large S) the ground state of this
model is the noncollinear $120^\circ$ structure corresponding to a
commensurate spiral order.  For a quantum spin S=1/2 the interplay of
quantum fluctuations and frustration leads to anomalous ground state
properties and low energy excitations.  This has been studied for
considerable time by analytical \cite{jolicoeur:89,chubukov:94} as
well as numerical \cite{kawamura:84,bernu:94,white:07} methods.  It
was finally concluded that the magnetic order is stable even in the
quantum case, albeit with strongly reduced ordered (staggered) moment
$m_Q/(\mu_BS)\approx0.41$.  This reduction may quantitatively be
understood as the effect of zero-point fluctuations in linear spin
wave (LSW) theory.  It not only affects the moment size but also the
ground state energy, magnetization and homogeneous susceptibility.
These corrections are formally of the order (1/S).  It was found that
(1/S) corrections also strongly modify the LSW (classical) dispersion
itself \cite{starykh:06}, leading to an overall band width reduction,
and additional (roton-type) minimum and peculiar life-time effects
\cite {chernyshev:09}.  This agrees with the results of
high-temperature series expansions \cite{zheng:06c}.

While the theoretical picture is fairly well established, there is,
however, no good material candidate for the isotropic S=1/2 quantum
HAF on the triangular lattice.  The systems studied sofar have
considerable interlayer-exchange and in-plane anisotropies
\cite{nakatsuji:10}, in particular for the most well studied \CCC{}
\cite{coldea:02,coldea:03,zvyagin:14} and \CCB{}
\cite{ono:03,zvyagin:14} compounds and their substitutional series
\CCCB{} \cite{cong:11}.  Therefore it is not sufficient to just focus
on the isotropic case.  A related observation is made for the square
lattice (interaction-) frustrated $J_1$-J$_2$ HAF (see e.g.
Ref.~\onlinecite{schmidt:10} and references cited therein).  Most of
the layered vanadium oxide compounds corresponding to this model have
strongly different exchange constants with J$_2$ being actually
ferromagnetic while J$_1$ is antiferromagnetic.  This is not only a
superficial analogy because, as we discuss below the anisotropic
triangular HAF is equivalent to the square lattice J$_1$-J$_2$ HAF
model with one J$_2$ bond cut.  Therefore, throughout our work we will
emphasize similarities and differences of both models.

\begin{figure}
    \centering
    {\includegraphics[height=.3\columnwidth]{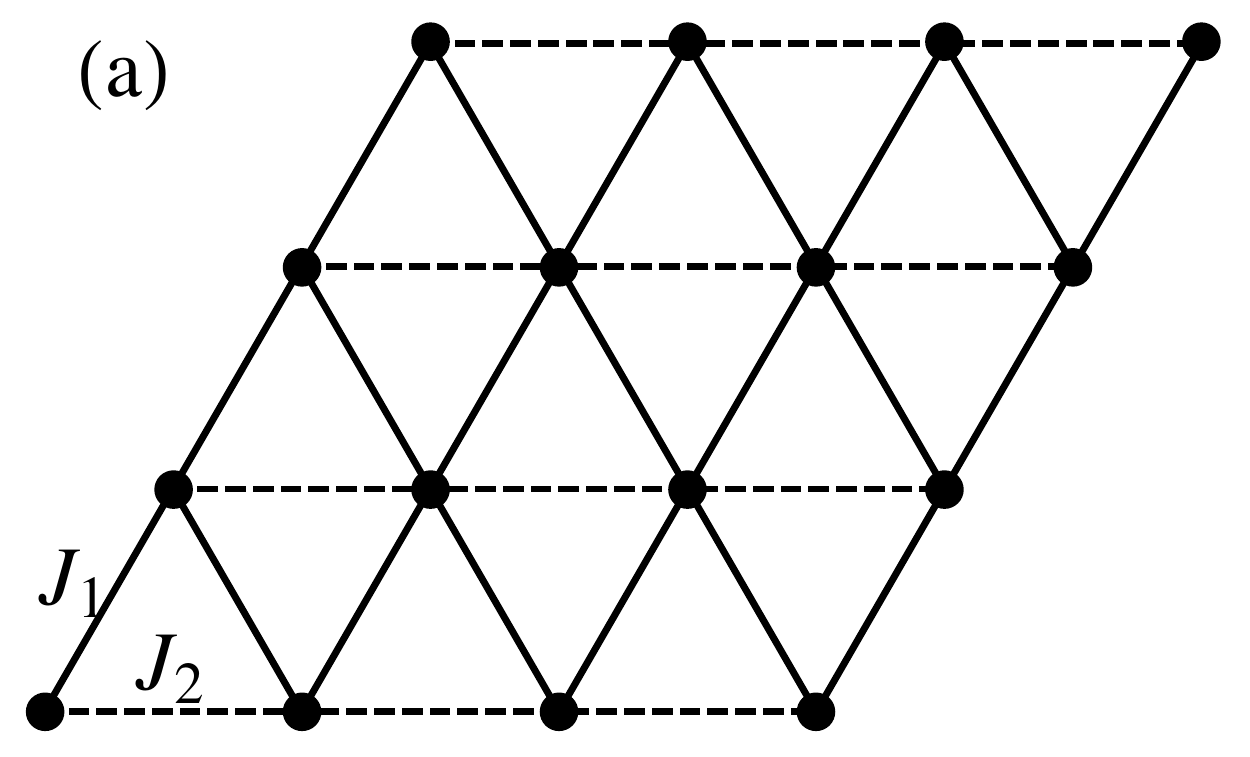}}
    \hspace{0.1cm}
    {\includegraphics[height=.3\columnwidth]{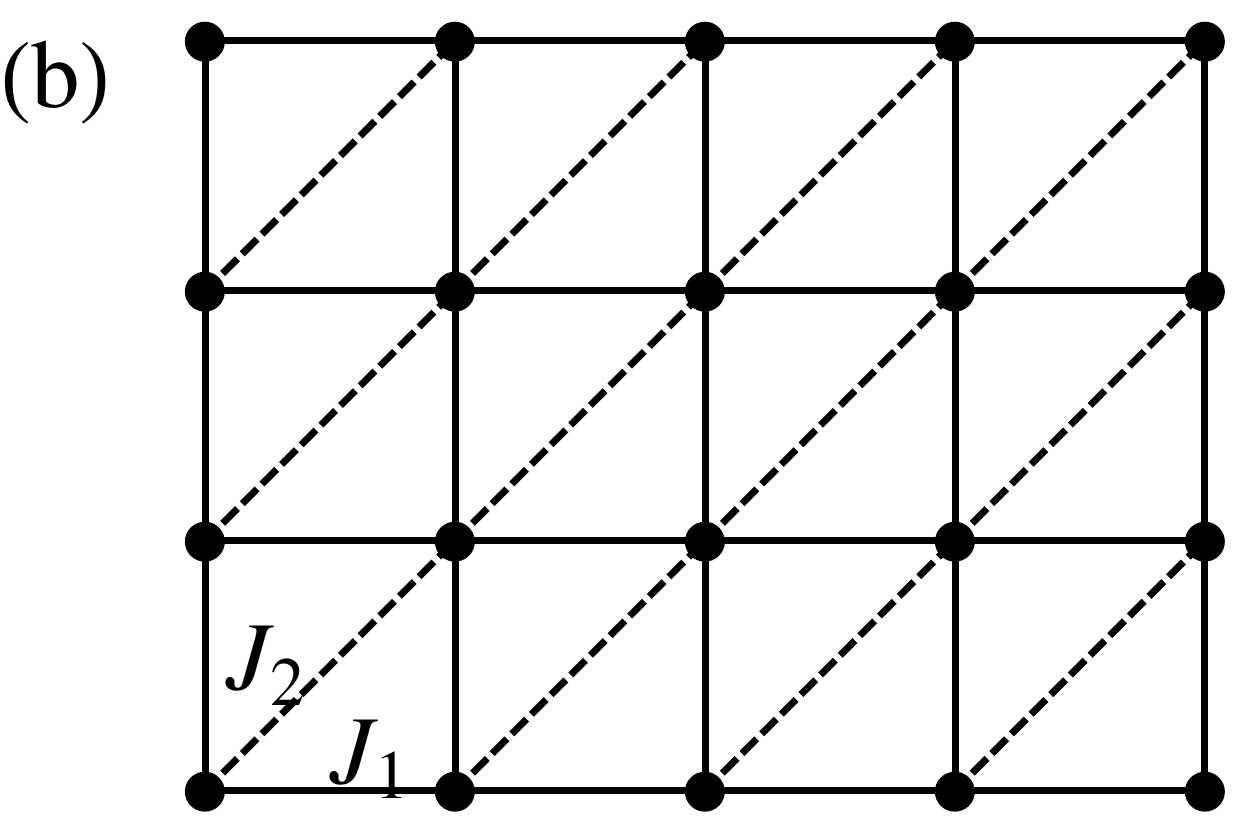}}
    \caption{(a) anisotropic triangular lattice with exchange
    constants $J_{1}$ (solid bonds) and $J_{2}$ (dashed bonds) which
    is topologically equivalent to a square lattice with one
    additional set of diagonal bonds (b).}
    \label{fig:geo}
\end{figure}
Due to this predominance of anisotropic exchange materials it is
highly useful not only to consider the isotropic triangular HAF but to
investigate the full phase diagram of the anisotropic model, in
particular because the substitutional series \CCCB{} may allow a
tuning of the anisotropy ratio within a certain range \cite{cong:11}.
It is a further interesting aspect of this model that by tuning a
single control parameter one may go from square lattice HAF
($\square$) via isotropic triangular ($\triangle$) to quasi-1D spin
chain ($\parallel$) system.  There exist several analytical
\cite{merino:99,trumper:99,veillette:05,veillette:05c,reuther:11} and
numerical \cite{weihong:99,weng:06,hauke:11,hauke:13} investigations
of the triangular model and its generalization \cite{doretto:12} but
only for the case when both anisotropic exchange constants J$_1$ and
J$_2$ are antiferromagnetic.

Here we present an analysis of the general triangular exchange model
without restriction to J$_1$ and J$_2$ (or the
anisotropy angle $\phi= \tan^{-1}(J_2/J_1)$). Furthermore we
investigate magnetization and field dependence of ordered moment in
the full
range of the anisotropy parameter. We use the linear spin
wave (LSW) theory in an external field as starting point of our
analysis, although one must keep in mind that this method is strictly
not applicable in the quasi-1D  and AF/spiral boundary regions of the
phase diagram.
Our emphasis is to derive the systematic trend of (1/S) quantum
corrections in ground state energy, ordering vector, staggered moment
and magnetization as function of the anisotropy angle
$\phi$ in the whole range $-\pi\leq \phi\leq\pi $. This should be
very useful information for the interpretation of real anisotropic
triangular magnets
and their location in the phase diagram as it has been for the square
lattice
cases, in particular for layered V oxides
\cite{shannon:04,tsirlin:09b} and also
for the quasi-1D system CsCuCl$_3$ with in-plane triangular structure
\cite{nikuni:93}.
Here we do not yet consider the (1/S) corrections to the excitation
spectrum itself as function of $\phi$, as was done for the isotropic
$\phi=\pi/4$ in Ref.\onlinecite{chernyshev:09}, this is left for a
later investigation.

\section{Model Hamiltonian}
\label{sec:hamiltonian}

The model spin Hamiltonian for the anisotropic triangular lattice is
given 
by
\begin{equation}
    {\cal H}
    =
    \sum_{\left\langle ij\right\rangle}\vec S_{i}J_{ij}\vec S_{j}
    -
    g\mu_{\text B}\mu_{0}\vec H\sum_{i}\vec S_{i}
    \label{eqn:h}
\end{equation}
where the sum in the first term extends over bonds $\langle ij\rangle$
connecting sites $i$ and $j$.  We assume an interaction in the form
of a uniaxial tensor in spin space defined by
\begin{equation}
    J_{ij}
    =
    \mathop{\rm diag}\left(
    J_{ij}^{\perp},J_{ij}^{\perp},J_{ij}^{z}
    \right),
\end{equation}
and the applied magnetic field $\vec H$ points into the $z$ direction
defined by the anisotropy above.  Figure~\ref{fig:geo}a illustrates
the spatial structure of $\cal H$, and
we set
\begin{equation}
    J_{ij}
    =
    \left\{
    \begin{array}{l@{\mbox{ if }}l}
        J_{1} & \vec R_{j}=
        \vec R_{i}\pm\frac 12\left(
        \vec e_{x}\pm\sqrt{3}\vec e_{y}\right)\\
        J_{2} & \vec R_{j}=
        \vec R_{i}\pm\vec e_{x}
    \end{array}
    \right.,
\end{equation}
measuring distances in units of the lattice constants.  For simplicity
we omit the directional indices $\perp$ and $z$ in the following,
unless otherwise
noted.

\begin{figure}
    \centering
    \includegraphics[width=.6\columnwidth]{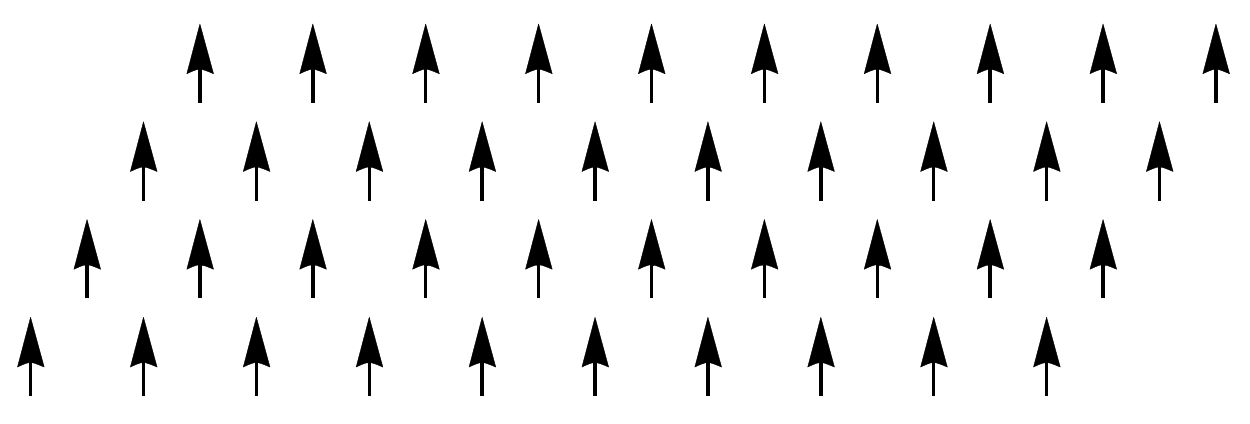}
    \includegraphics[width=.6\columnwidth]{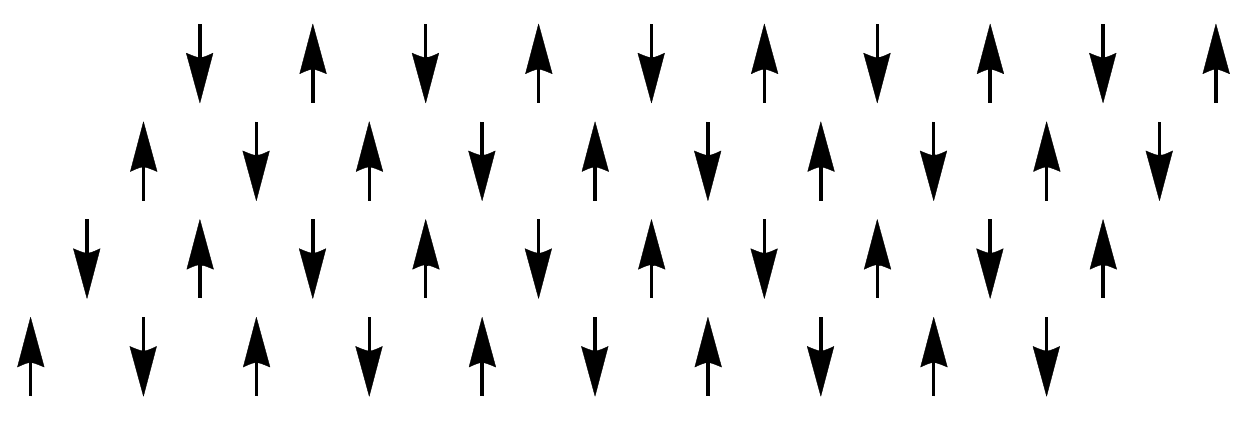}
    \includegraphics[width=.6\columnwidth]{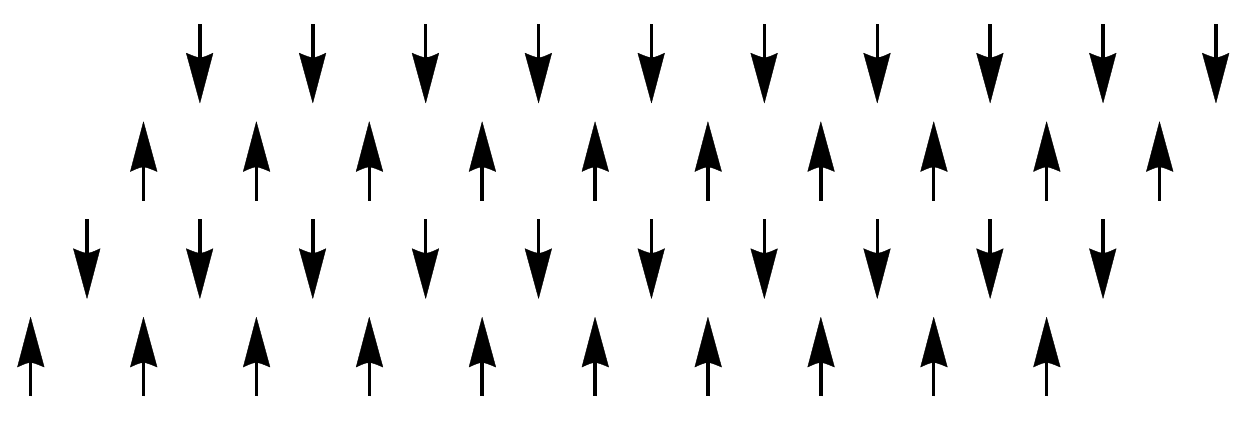}
    \includegraphics[width=.6\columnwidth]{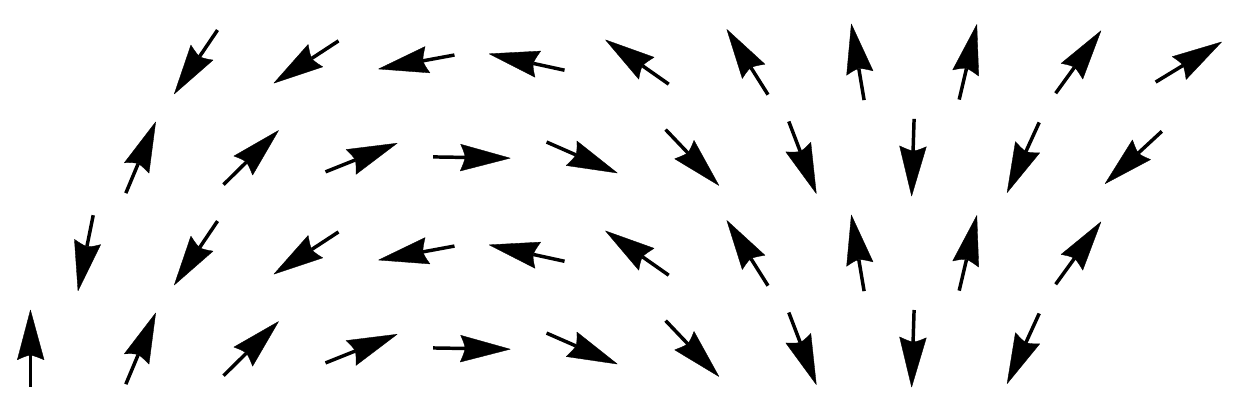}
    \includegraphics[width=.6\columnwidth]{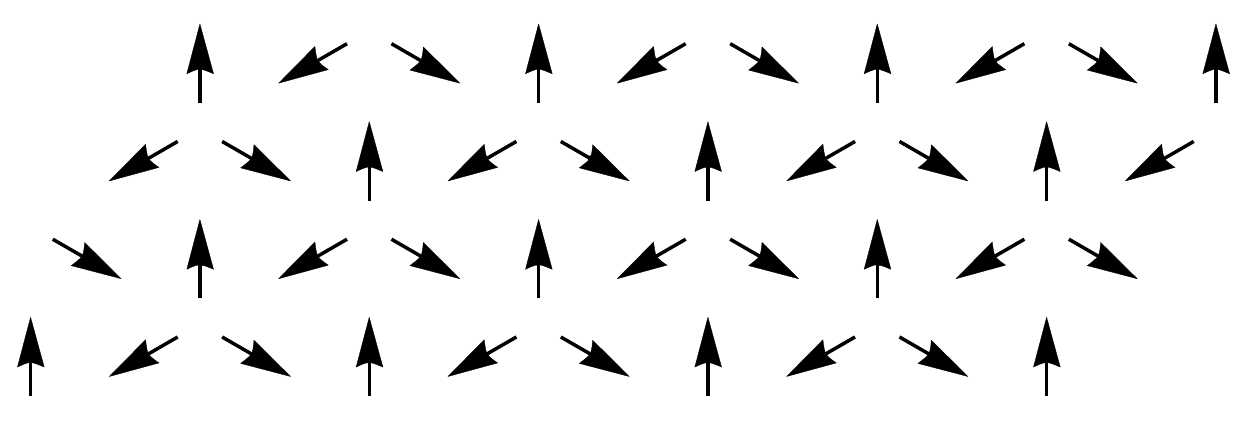}
    \caption{Moment patterns of the ordered states, from top to
    bottom: Ferromagnet, columnar antiferromagnet, antiferromagnet,
    spiral with $J_{2}/J_{1}=1/2+0.01$, special case:
    120-degree-structure for $J_{2}=J_{1}>0$ (isotropic AF exchange).}
    \label{fig:pat}
\end{figure}

The Fourier transform of the exchange integral for the lattice with 
$N$ sites then reads
\begin{eqnarray}
    J(\vec q)
    &=&
    \frac 1N\sum_{\langle ij\rangle}J_{ij}
    {\rm e}^{-{\rm i}\vec q\left(\vec R_{i}-\vec R_{j}\right)}
    =
    \frac 12\sum_{n}J_{ii+n}{\rm e}^{-{\rm i}\vec q\vec R_{n}}
    \nonumber\\
    &=&
    2J_{1}\cos\frac{q_{x}}2\cos\frac{\sqrt{3}}2q_{y}
    +
    J_{2}\cos q_{x},
\end{eqnarray}
where the last sum runs over all bonds $n$ connecting an arbitrary but
fixed site $i$ with its six neighbors. Alternatively, we can regard 
the lattice as an isotropic square lattice with one additional 
diagonal bond as shown in  Fig.~\ref{fig:geo}b. 
The mapping corresponds to a coordinate transformation in crystal 
momentum space
\begin{equation}
    q_{x}\to k_{x}+k_{y},
    \quad
    q_{y}\to \frac 1{\sqrt{3}}\left(k_{x}-k_{y}\right),
\end{equation}
and the exchange Fourier transform acquires the form
\begin{equation}
    J(\vec k)
    =
    J_{1}\left(\cos k_{x}+\cos k_{y}\right)
    +J_{2}\cos\left(k_{x}+k_{y}\right).
\end{equation}
We will use the symbols $q,Q$ for the triangular lattice coordinates,
and use $k,K$ for the square lattice coordinates.

In order to discuss the full phase diagram of the model, we introduce
a anisotropy angle $\phi$ and an overall energy scale $J_{\text c}$
defined through
\begin{eqnarray}
    J_{1}=J_{\text c}\cos\phi,
    &\quad&
    J_{2}=J_{\text c}\sin\phi,
    \\
    J_{c}=\sqrt{J_{1}^{2}+J_{2}^{2}},
    &\quad&
    \phi=\tan^{-1}\left(\frac{J_{2}}{J_{1}}\right).
    \nonumber
\end{eqnarray}
This parameterization allows for an interpolation between important
geometrical limiting cases, namely the square-lattice Néel
antiferromagnet ($J_{2}=0$ or $\phi=0$), the isotropic triangular
antiferromagnet ($J_{2}=J_{1}$ or $\phi=\pi/4$), the antiferromagnetic
chain ($J_{1}=0$ or $\phi=\pi/2$), and their ferromagnetic
counterparts.
The following analysis of the model with the parameterization
introduced here closely follows the general concept presented
in~\cite{schmidt:10}.

\begin{figure}
    \centering
    \includegraphics[width=.9\columnwidth]{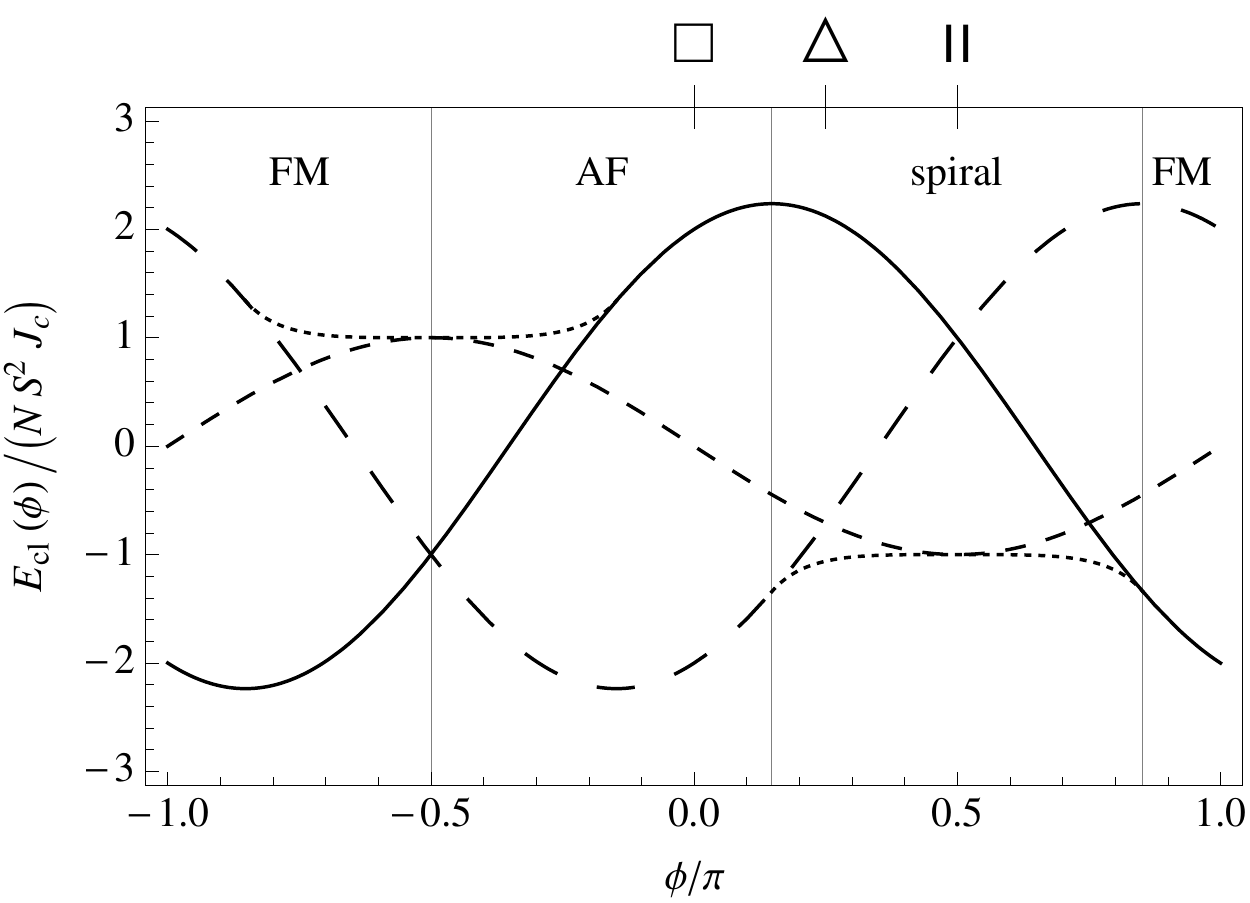}
    \includegraphics[width=.9\columnwidth]{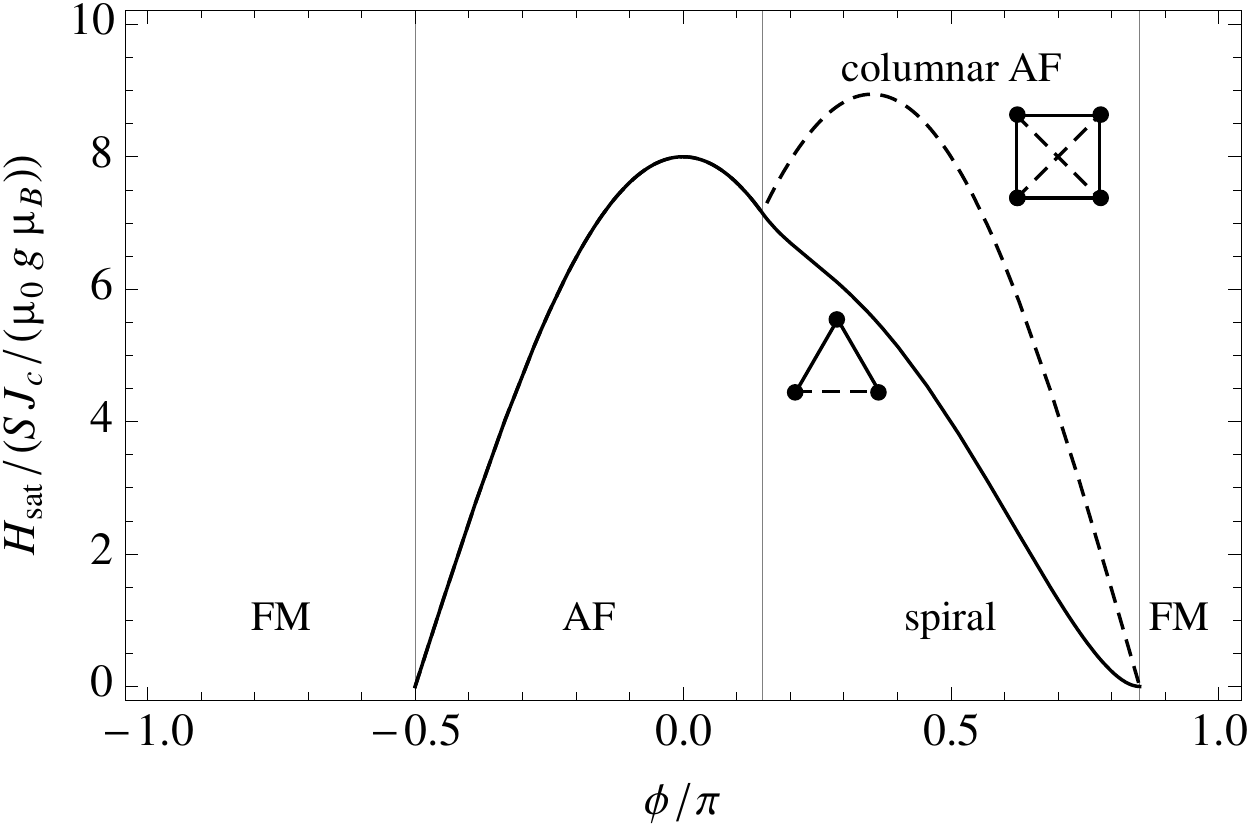}  
    \caption{Top: Dependence of the classical energies listed in
    Table~\ref{tbl:ecl} on the anisotropy angle $\phi$.  Solid line:
    ferromagnet, long-dashed line: antiferromagnet, dashed line:
    columnar antiferromagnet, dotted line: spiral.  The vertical lines
    mark the phase boundaries, the special anisotropy angles for the
    antiferromagnetic square lattice ($J_{2}=0$, $\Box$), the
    isotropic triangular antiferromagnet ($J_{2}=J_{1}$, $\triangle$),
    and the antiferromagnetic chains ($J_{1}=0$, $\|$) are indicated
    by the corresponding symbols above the plot.  Bottom: Saturation
    field for triangular (solid line) and square lattice $J_1-J_2$
    model (dashed line).}
    \label{fig:ecl}
\end{figure}
\section{Classical ground states}
\label{sec:classical}

On each site $i$, we introduce a local coordinate system where the
local $z'$ axis is oriented parallel to the local magnetic moment,
which we regard as a classical vector.  The global $z$ axis is defined
by the direction of the magnetic field $\vec H$.  Ignoring
anisotropies, the spins undergo a spin-flop
transition for arbitrary small fields.  We can parameterize them with
a canting angle $\Theta$
relative to the field direction and an ordering vector $\vec Q$ in the
$xy$ plane perpendicular to the field direction.  The value
$\Theta=\pi/2$ corresponds to the state with vanishing magnetic field
(arbitrary global $z$ axis), and $\Theta=0$ corresponds to the fully 
polarized state at the saturation field.

\begin{table*}
    \caption{Classical ground states: Ordering vectors, energies,
    conditions, range of anisotropy angle.  For each phase the
    possible equivalent wave vectors in the first BZ that produce the
    same spin structure are given.  }\vspace{0.3cm}
    \begin{ruledtabular}
    \begin{tabular}{lllll}
        phase
        &
        classical ordering vectors $\vec Q$
        &
        $E_{\text{cl}}/(NS^{2})$
        &
        conditions
        &
        range
        \\
        \hline
        ferromagnet
        &
        $
        \begin{array}{l}
            0
            \\
            \left(\pm2\pi,\pm\frac{2\pi}{\sqrt{3}}\right)
        \end{array}
        $
        &
        $2J_{1}+J_{2}$
        &
        $J_{1}\le0\ \wedge\ \frac{J_{2}}{|J_{1}|}\le\frac 12$
        &
        $-\pi-\tan^{-1}\left(\frac 12\right)\le\phi\le-\frac{\pi}2$
        \\
        \hline
        columnar AF
        &
        $\left(\pm\pi,\pm\frac{\pi}{\sqrt{3}}\right)$
        &
        $-J_{2}$
        &
        --
        &
        --
        \\
        \hline
        antiferromagnet
        &
        $
        \begin{array}{l}
            \left(\pm2\pi,0\right)
            \\
            \left(0,\pm\frac{2\pi}{\sqrt{3}}\right)
        \end{array}
        $
        &
        $-2J_{1}+J_{2}$
        &
        $J_{1}\ge0\ \wedge\ \frac{J_{2}}{J_{1}}\le\frac 12$
        &
        $-\frac{\pi}2\le\phi\le\tan^{-1}\left(\frac 12\right)$
        \\
        \hline
        spiral
        &
        $
        \begin{array}{l}
            \left(
            \pm2\tan^{-1}\left(
            \sqrt{4\left(\frac{J_{2}}{J_{1}}\right)^{2}-1}
%             \frac{\sqrt{4J_{2}^{2}-J_{1}^{2}}}{J_{1}}
            \right),
            0
            \right)
            \\
            \left(
            \pm2\tan^{-1}\left(
            \sqrt{4\left(\frac{J_{2}}{J_{1}}\right)^{2}-1}
%             \frac{\sqrt{4J_{2}^{2}-J_{1}^{2}}}{J_{1}}
            \right),
            \pm\frac{2\pi}{\sqrt{3}}
            \right)
        \end{array}
        $
        &

$-J_{2}\left[1+\frac12\left(\frac{J_{1}}{J_{2}}\right)^{2}\right]$
%         $-J_{2}\left(1+\frac{J_{1}^{2}}{2J_{2}^{2}}\right)$
        &
        $J_{2}\ge0\ \wedge\ \frac{J_{2}}{|J_{1}|}\ge\frac 12$
        &
        $\tan^{-1}\left(\frac 12\right)\le\phi\le
        \pi-\tan^{-1}\left(\frac 12\right)$
	\\
	\hline
	isotropic triangular AF
	&
	$
	\begin{array}{l}
            \left(\pm\frac{4\pi}{3},0\right)
            \\
             \left(\pm\frac{2\pi}{3},\pm\frac{2\pi}{\sqrt{3}}\right)
       \end{array}
        $
	&
	$-\frac32J_{1}$
	&
	$J_{1}=J_{2}>0$
	&
	$\phi=\frac\pi4$
    \end{tabular}
    \end{ruledtabular}
    \protect\label{tbl:ecl}
\end{table*}
The classical energy then reads
\begin{equation}
    E_{\text{cl}}
    =
    NS^{2}\left[
    J_{\perp}(\vec Q)+A(0)\cos^{2}\Theta
    -\frac{g\mu_{\text B}\mu_{0}H}S\cos\Theta
    \right],
    \label{eqn:ecl}
\end{equation}
where the coefficient $A(0)=A(\vec q=0)$ is defined by
\begin{equation}
    A(\vec q)
    =
    J_{z}(\vec q)+\frac 12\left[
    J_{\perp}(\vec q+\vec Q)+J_{\perp}(\vec q-\vec Q)
    \right]
    -2J_{\perp}(\vec Q).
    \label{eqn:a}
\end{equation}
Minimizing Eq.~(\ref{eqn:ecl}) with respect to $\Theta$ yields the
classical canting angle and saturation field $H_{\text{sat}}$:
\begin{equation}
    \Theta_{\text c}
    =
    \cos^{-1}\left(
    \frac H{H_{\text{sat}}}
    \right),
    \quad
    H_{\text{sat}}=\frac{2SA(0)}{\mu_{0}g\mu_{\text B}},
\end{equation}
and we get
\begin{equation}
    E_{\text{cl}}
    =
    NS^{2}\left[
    J_{\perp}(\vec Q)-A(0)
    \left( \frac H{H_{\text{sat}}}\right)^{2}
    \right].
\end{equation}
The vector $\vec Q$ can be found by minimizing the above expression
with respect to it.  Setting $\partial E_{\text{cl}}/\partial
Q_{\alpha}=0$, $\alpha=x,y$ we find different types of solutions which
are tabulated in Table~\ref{tbl:ecl}, together with their classical
energies.  The last two columns of the table list the conditions and
ranges of anisotropy angle where the corresponding solution $\vec Q$
describes the ground state.

Fig.~\ref{fig:pat} illustrates the moment patterns of the different
classical phases.  In the isotropic case, the columnar antiferromagnet
and the antiferromagnet are identical, because the former can be
mapped onto the latter through a rotation of the lattice by an angle
$2\pi/3$.  Therefore their energy curves cross at $\phi=\pi/4$
(Fig~\ref{fig:ecl} (top)).  However the ordering vector corresponding
to the columnar antiferromagnet never minimizes the classical energy,
and for $J_{1}=J_{2}>0$ ($\phi=\pi/4$), the 120-degree-structure
characterizes the ground state (last row in Table~\ref{tbl:ecl}).
Figure~\ref{fig:ecl} (top) displays the dependence of the classical
energies on the anisotropy angle $\phi$ while Figure~\ref{fig:ecl}
(bottom) shows the saturation fields in comparison of triangular and
square lattice $J_1-J_2$ model.

\begin{figure}
    \centering
    \includegraphics[width=.8\columnwidth]{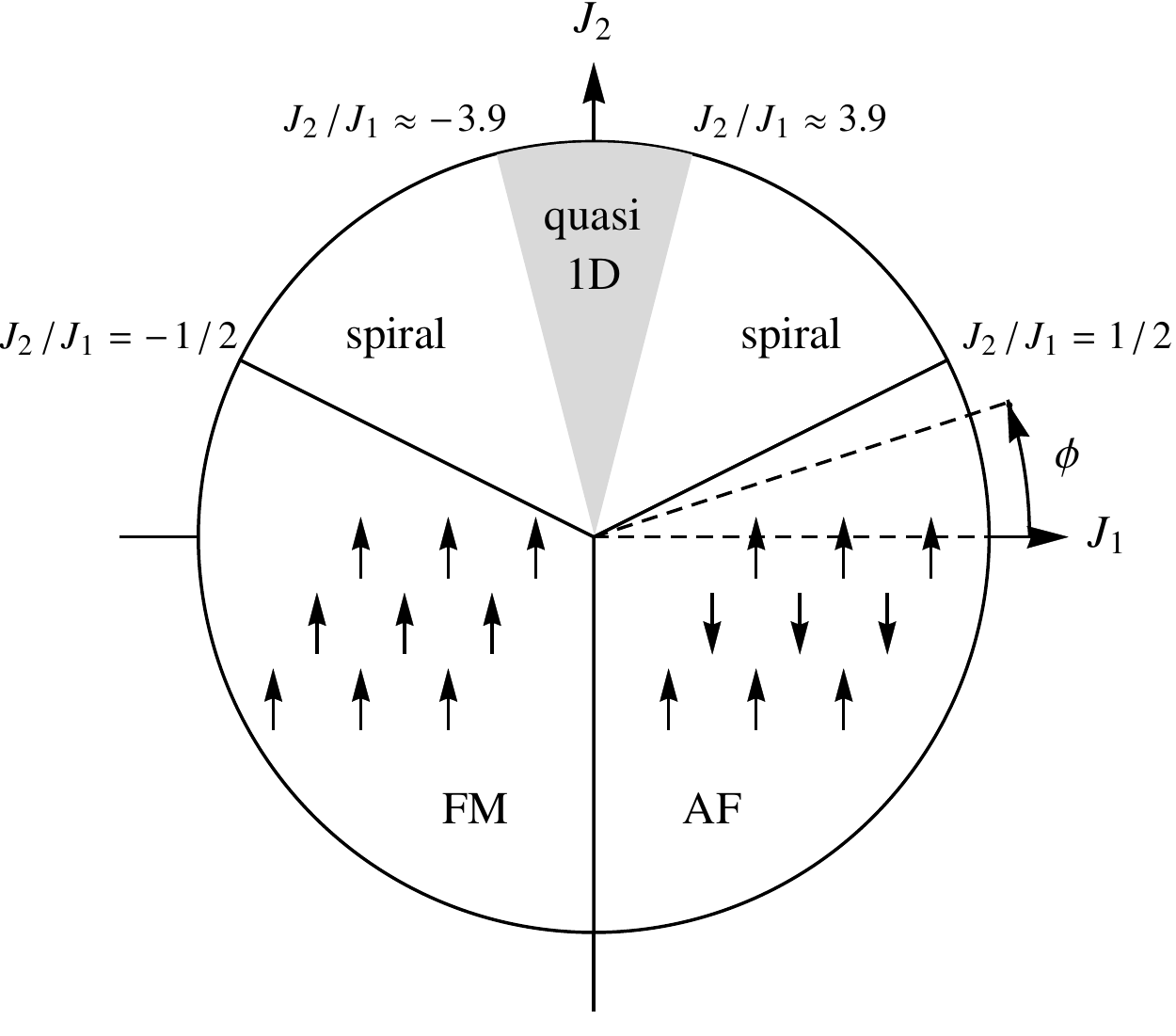}
    \caption{Classical phase diagram. 
    The phase boundaries are
    determined from the classical ground-state energy.  The boundaries
    of the spiral phase are given by $J_{2}/J_{1}=\pm1/2$, the
    boundary between AF and FM is given by $J_{1}=0$, $J_{2}\ge0$.
    The shaded area bounded by $J_{2}/J_{1}\approx\pm3.9$
    ($0.42\pi\le\phi\le0.58\pi$) denotes the region where quantum
    fluctuations destroy the ordered moment within our semiclassical 
    approximation. This area is centered around $J_{1}=0$, which 
    corresponds to one-dimensional chains. The dashed lines 
    illustrate the definition of the anisotropy angle $\phi$ with
$-\pi\leq\phi\leq\pi$.}
    \label{fig:pd}
\end{figure}
The phase diagram according to this analysis is composed of three
phases, a uniform ferromagnet (FM), a Néel antiferromagnet (AF), and a
spiral phase with an in general incommensurate ordering vector, see
Fig.~\ref{fig:pd}.  The columnar antiferromagnet has an energy which
is never the lowest, therefore it is not realized as a ground state.

\begin{figure}
    \centering
    \includegraphics[width=.9\columnwidth]{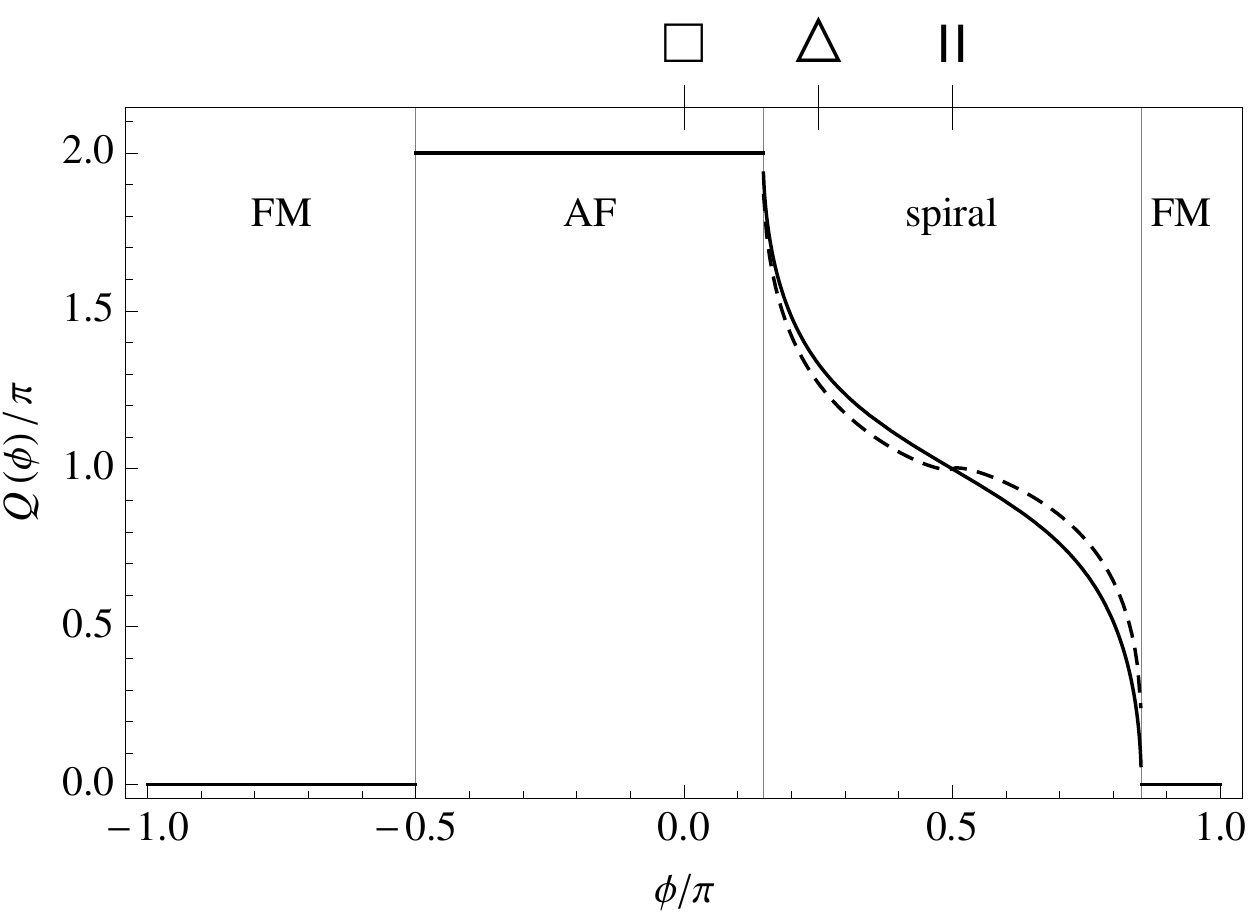}
    \caption{Dependence of the component $Q$ of the ordering vector
    $\vec Q=(\pm Q,0)$ on the anisotropy angle $\phi$. The solid 
    line denotes the classical behavior, the dashed line denotes the 
    $\phi$-dependence including first-order quantum corrections.}
    \label{fig:q}
\end{figure}

One set of ordering vectors has the form $\vec Q=(\pm Q,0)$ for all
three classical phases, shown in Fig.~\ref{fig:q}.  In the spiral
phase, $Q$ continuously interpolates between the antiferromagnet
($Q=2\pi$) and the ferromagnet ($Q=0$).  In this phase,
$Q(\phi)=2\pi-Q(\pi-\phi)$ is antisymmetric with respect to the point
$(\phi,Q)=(\pi/2,\pi)$ (antiferromagnetic chain with $J_{1}=0$,
$J_{2}>0$) and has an infinite slope at the boundaries.  The isotropic
triangular lattice with $J_{1}=J_{2}$ is characterized by
$Q(\pi/2)=4\pi/3$ leading to the $120^\circ$ spin structure.

The phase diagram has a corresponding mirror symmetry with respect to
the line $J_{1}=0$: We may split the lattice into a sublattice $A$
having only sites with a particular spin direction, e.g. ``up'', in
the AF phase, and a sublattice $B$ with sites all having a ``down''
spin.  If all $B$ spins are flipped the Hamiltonian~(\ref{eqn:h}) with
$\vec H=0$ remains invariant provided we simultaneously set
$J_{1}\to-J_{1}$.  This operation exchanges the FM and AF phases and
leaves the spiral phase invariant.  Exactly at $J_{1}=0$, the lattice
is decoupled into independent chains, and one can rotate all spins on
any chain by an arbitrary angle without energy cost.

\section{Quantum fluctuations}
\label{sec:fluctuations}

The classical picture of magnetic order in the anisotropic trigonal
model discussed sofar serves as a starting point to the main topic of
this work: A global understanding of the influence of quantum
fluctuations on the type, energy, ordered moment and magnetization of
the ground state as function of the anisotropy angle $\phi$.  For that
purpose we employ a standard linear spin wave analysis of magnetic
excitations and calculate the effect of zero point fluctuations on
these quantities.  Naturally, as in the square lattice model, regions
of $\phi$ where the moment becomes unstable and spin wave
approximation breaks down will appear.  It is important to
characterize the size of those regions and get clues on the type of
nonmagnetic state that may appear.

\begin{figure}
    \centering
    \includegraphics[width=.9\columnwidth]{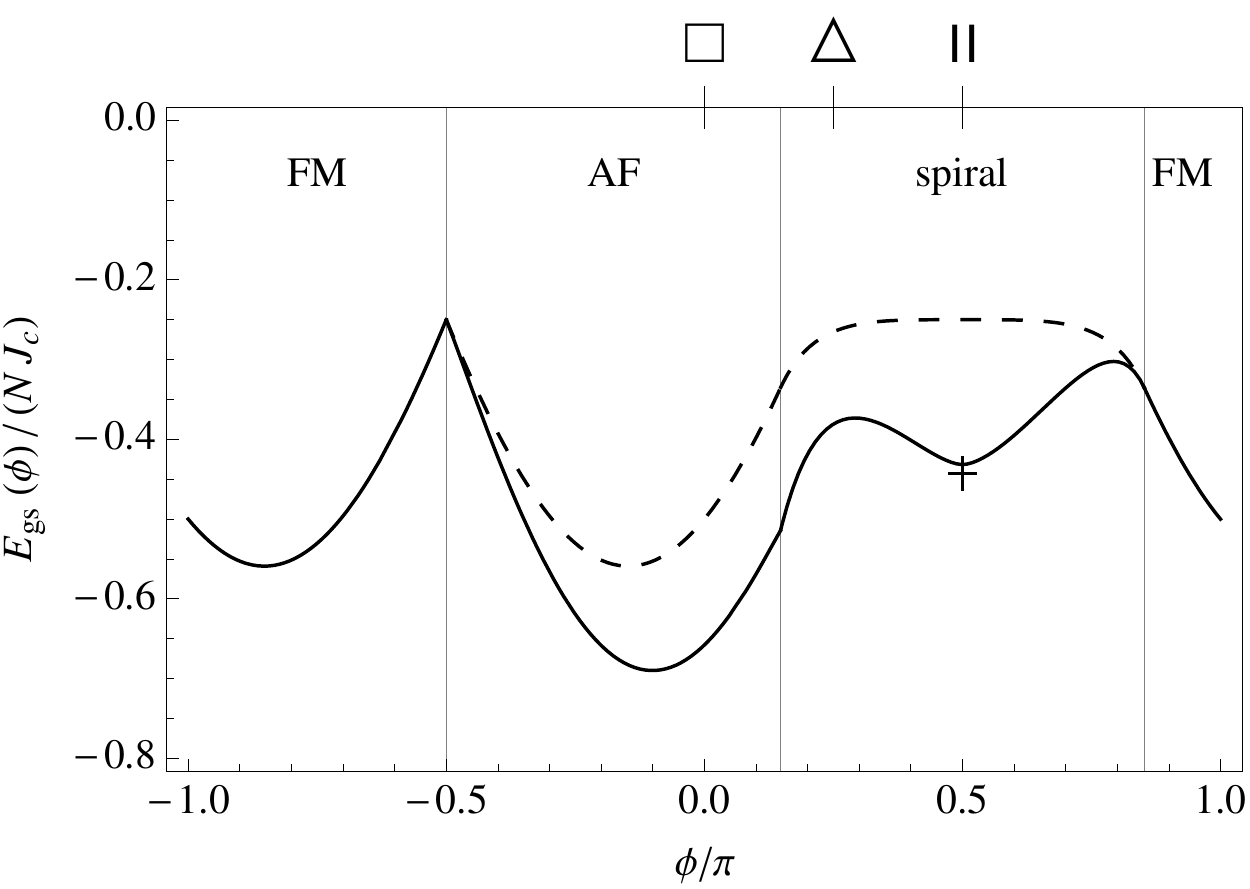}

    \includegraphics[width=.9\columnwidth]{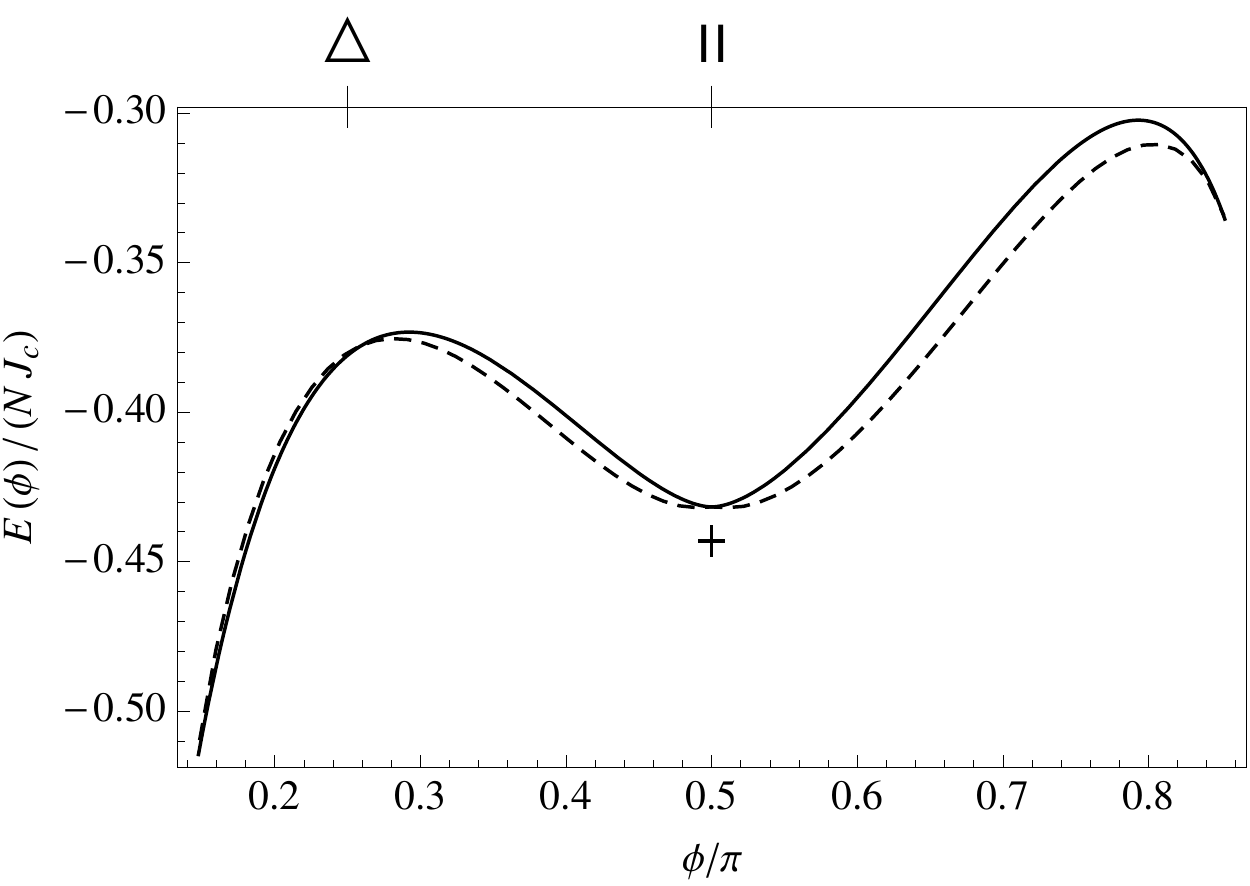}
    \caption{Ground-state energy $E_{\text{gs}}$ for $S=1/2$ including
    first-order quantum corrections as a function of $\phi$ (solid
    line).  Cross at $\phi=\pi/2$: Exact 1D Bethe ansatz result
    $E_{\text{gs}}/(NJ_{\text c})=1/4-\ln2$.  Top: Full range of
    anisotropy angles.  Dashed line: Classical ground-state energy.
    Bottom: Zoom into the spiral phase.  Dashed line: small
    corrections to $E_{\text{gs}}$ in the spiral phase due to ordering
    vector corrections.}
    \label{fig:egs}
\end{figure}
To perform the calculation of quantum corrections we return to the
Hamiltonian~(\ref{eqn:h}) expressed in the local coordinate system
introduced in the previous section, we apply a Holstein-Primakoff
transformation and carry out a large-$S$ expansion, keeping terms up
to fist order in $1/S$.  We regard the magnetic field $\vec H$
formally as
proportional to $S$.  Applying a Fourier transformation and a
subsequent Bogoliubov transformation leads in the thermodynamic limit
to the familiar general form
\begin{equation}
    {\cal H}
    =
    E_{\text{cl}}
    +
    E_{\text{zp}}
    +
    NS\int\frac{{\rm d}^{2}q}{V_{\text{BZ}}}
    \omega_{\text{sw}}(\vec q)
    \alpha_{\vec q}^{\dagger}\alpha_{\vec q},
\end{equation}
where the $q$ integration is to be taken over the first Brillouin
zone (BZ) with area $V_{\text{BZ}}=2\sqrt3\pi^{2}$, $E_{\text{cl}}$ is
given by Eq.~(\ref{eqn:ecl}), and $\alpha^{\dagger}_{\vec q}$ denotes
a magnon with wave vector $\vec q$.  The zero-point energy
$E_{\text{zp}}$ and magnon dispersion $\omega_{\text{sw}}$ are given
by
\begin{eqnarray}
    E_{\text{zp}}
    &=&
    NS\left(J_{\perp}(\vec Q)
    +
    \frac 12\int\frac{{\rm d}^{2}q}{V_{\text{BZ}}}\omega(\vec q)
    \right),
    \label{eqn:ezp}
    \\
    \omega_{\text{sw}}(\vec q)
    &=&
    \omega(\vec q)+C(\vec q)\frac H{H_{\text{sat}}},
    \\
    \omega(\vec q)
    &=&
    \left\{\left[
    A(\vec q)-B(\vec q)\left(H/H_{\text{sat}}\right)^{2}
    \right]^{2}
    \right.
    \nonumber\\&&
    \left.{}
    -\left[
    B(\vec q)\left(1-\left(H/H_{\text{sat}}\right)^{2}\right)
    \right]^{2}\right\}^{1/2}.
    \label{eqn:omega}
\end{eqnarray}
The coefficients $B$ and $C$, the latter only being present at finite
magnetic fields $\vec H$, are given by
\begin{eqnarray}
    B(\vec q)
    &=&
    J_{z}(\vec q)-\frac 12\left[
    J_{\perp}(\vec q+\vec Q)+J_{\perp}(\vec q-\vec Q)
    \right],
    \label{eqn:b}
    \\
    C(\vec q)
    &=&
     J_{\perp}(\vec q+\vec Q)-J_{\perp}(\vec q-\vec Q).
\end{eqnarray}
We note that $C(\vec q)$ is nonzero only in the spiral phase and does
not contribute to the ground-state energy $E_{\text{gs}}=E_{\text{cl}}
+ E_{\text{zp}}$.

For the determination of the ordering vector $\vec Q$, we now have to
minimize the full ground-state energy $E_{\text{gs}}=E_{\text{cl}} +
E_{\text{zp}}$ as given by Eqs.~(\ref{eqn:ecl}) and~(\ref{eqn:ezp}),
again up to first order in $1/S$.  For $\vec H=0$, this amounts to
finding the roots of the equation
\begin{equation}
    -\frac{\partial J_{\perp}(\vec Q)}{\partial Q_{\alpha}}
    =
    \frac{1}{2S}\int\frac{{\rm d}^{2}q}{V_{\text{BZ}}}\left.\left(
    \frac{\partial A(\vec q)}{\partial Q_{\alpha}}
    \cdot
    \frac{A(\vec q)+B(\vec q)}{\omega(\vec q)}
    \right)\right|_{\vec Q=\vec Q_{\text{cl}}}
	\label{eqn:min}
\end{equation}
for the components of $\vec Q$.  In the AF and FM phases, the
right-hand side of this equation vanishes, and Eq.~(\ref{eqn:min})
leads again to the classical values a given in Table~\ref{tbl:ecl}.
Small but finite corrections to these appear in the spiral phase only,
see the dashed line of Fig.~\ref{fig:q}.  Unless explicitly noted, we
ignore these as well as spin-space anisotropies
($J_{ij}^{\perp}=J_{ij}^{z}$) in the following.

\begin{figure}
    \centering
    \includegraphics[width=.9\columnwidth]{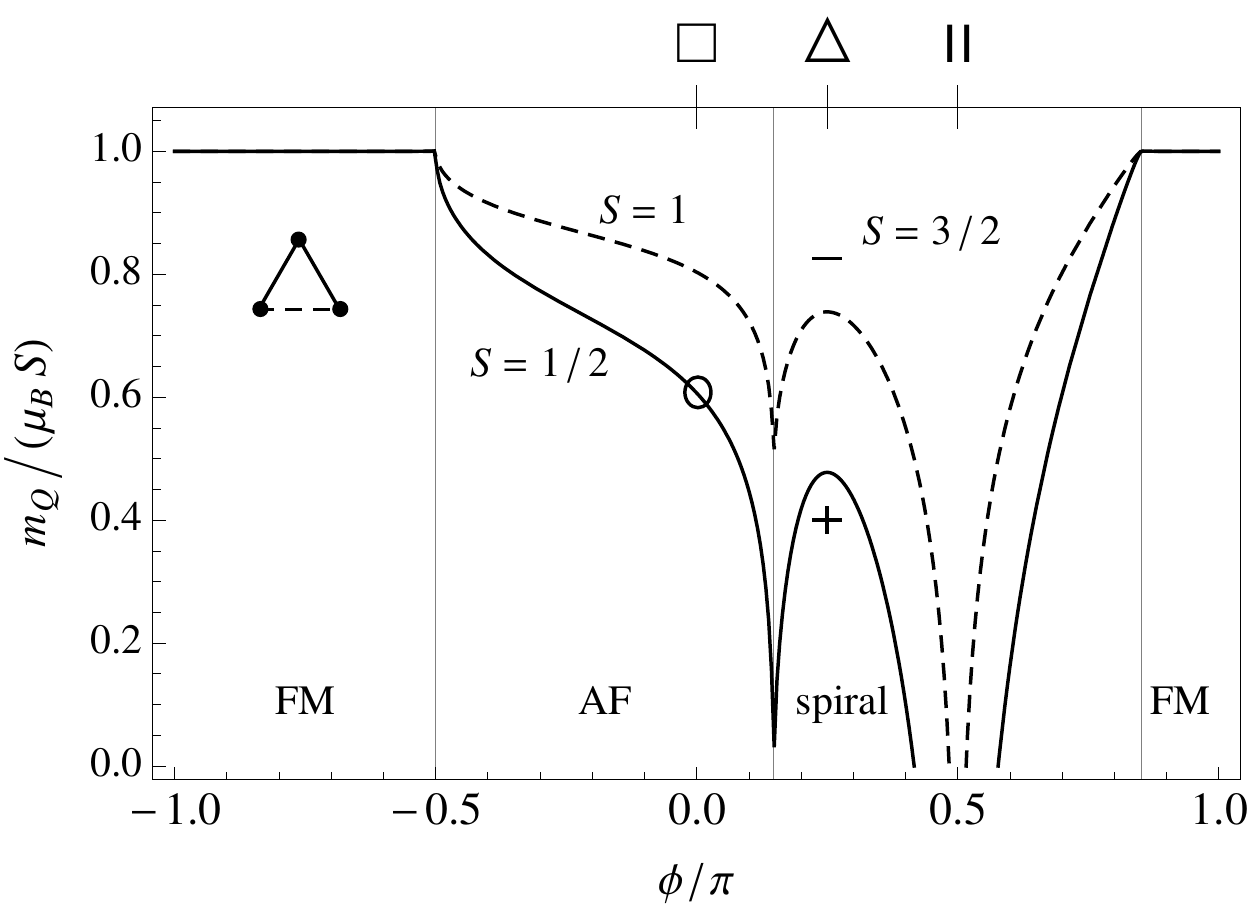}
    \includegraphics[width=.9\columnwidth]{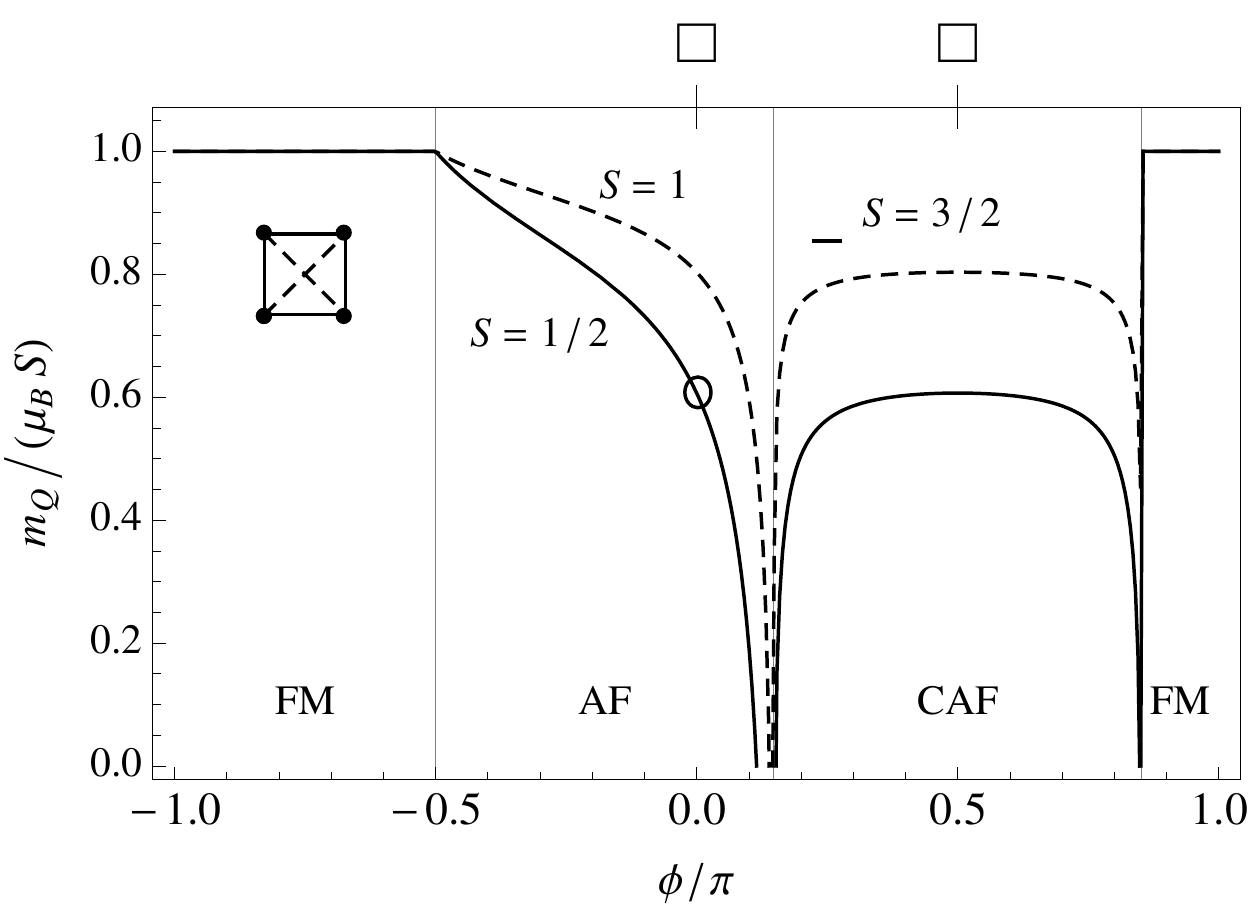}
    \caption{Size of the ordered moment $m_{Q}$ according to
    Eq.~(\ref{eq:ordered}) as a function of the anisotropy angle
    $\phi$.  Top: anisotropic triangular lattice.  Bottom:
    $J_{1}$-$J_{2}$ model on the square lattice.  The solid lines
    denote $S=1/2$, the circle marks the ordered moment for the
    square-lattice antiferromagnet.  The cross marks the DMRG result
    $m_Q/(\mu_{\text B}S)=0.41$~\cite{white:07}.  The dotted lines
    show the $\phi$ dependence of $m_{Q}$ for $S=1$, and the bar at
    $\phi=\pi/4$ denotes $m_{Q}$ for $S=3/2$.}
    \label{fig:mo}
\end{figure}

\section{Ground-state energy}
\label{sec:groundstate}

In the ferromagnetic phase, corrections to the ground-state energy are
absent.  Quantum fluctuations strongly renormalize the ground-state
energy $E_{\text{gs}}$ in the antiferromagnetic and spiral phases.
The solid line in Fig.~\ref{fig:egs} shows a plot of $E_{\text{gs}}$
for $S=1/2$ as introduced in the preceding section, calculated using
the classical ordering vector.  For comparison, the dashed line in the
upper part of the figure displays the classical ground-state energy
$E_{\text{cl}}$.  The reflection symmetry of the classical energy
around the one-dimensional point ($\parallel$) is destroyed by the
quantum fluctuations.

While vanishing at the boundaries to the ferromagnet, the quantum
corrections are largest in the spiral phase, in particular around the
one-dimensional point $\phi=\pi/2$ ($J_{1}=0$).  The lower part of
Fig.~\ref{fig:egs} gives a zoom into this phase and shows the
additional small corrections due to the ordering vector corrections
introduced in the preceding section.  The cross at $\phi=\pi/2$
denotes the exact Bethe ansatz result $E_{\text{gs}}/(NJ_{\text
c})=1/4-\ln2\approx-0.443$ for the one-dimensional AF
chain~\cite{bethe:31,hulthen:38}.  From linear spin-wave theory we get
$E_{\text{gs}}/(NJ_{\text c})=-3/4+1/\pi\approx-0.432$.

\section{Ordered moment}
\label{sec:moment}

Quantum fluctuations reduce the ordered moment $m_{Q}$ from its
classical value $m_{Q}^{\text{cl}}=\mu_{\text B}S$.  Including 
their effect it has the general form
\begin{equation}
    m_{Q}
    =
    \mu_{\text B}S\left[
    1-\frac{1}{2S}\left(
    \int\frac{{\rm d}^{2}q}{V_{\text{BZ}}}
    \frac{A(\vec q)}{\omega(\vec q)}
    -1
    \right)
    \right]
    \label{eq:ordered}
\end{equation}
for field $\vec H=0$.  Fig.~\ref{fig:mo} shows a plot of the ordered
moment $m_{Q}$ at field $\vec H=0$ as a function of the anisotropy
angle $\phi$.  We note that the absolute correction to the classical
value $m_{Q}^{\text{cl}}$ does not depend on $S$.  In the FM phase,
the classical value remains unchanged.  In the AF phase, quantum
fluctuations reduce the ordered moment smoothly form a value
$m_{Q}=\mu_{\text B}S$ in the FM phase to $m_{Q}=0$ at
$J_{2}/J_{1}=1/2$ ($\phi\approx0.148\pi$).  At $\phi=0$, we get the
well-known value $m_{Q}/(\mu_{\text B}S)\approx0.606$ for $S=1/2$
(small circle in Fig.~\ref{fig:mo}).

In the spiral phase for $\phi\ge\tan^{-1}(1/2)$, the ordered moment is
stabilized until the commensurate $120^{\circ}$-spiral state with
ordering vector component $Q=4\pi/3$ is reached at $\phi=\pi/4$
(isotropic case), and $m_{Q}/(\mu_{\text B}S)\approx0.478$ for
$S=1/2$~\cite{chubukov:94}.  Increasing the ratio $J_{2}/J_{1}$
further, the ordered moment is suppressed until it vanishes at
$\phi/\pi\approx0.42$, corresponding to $J_{2}/J_{1}\approx3.9$.  For
$\phi/\pi>0.58$ or $J_{2}/J_{1}>-3.9$, the ordered moment is
stabilized and monotonously increases until it reaches the border to
the FM phase at $J_{2}/J_{1}=-1/2$ ($\phi/\pi\approx0.852$), where the
classical value is restored.  The region with $J_{2}/|J_{1}|>3.9$
indicates a classically not present disordered regime around the point
$\phi=\pi/2$, where $J_{1}=0$.  This point denotes the one-dimensional
antiferromagnetic chain which has no long range magnetic order
but only quasi-long range algebraic spin correlations
\cite{affleck:98}.

\begin{figure}
    \centering

    \includegraphics[width=.9\columnwidth]{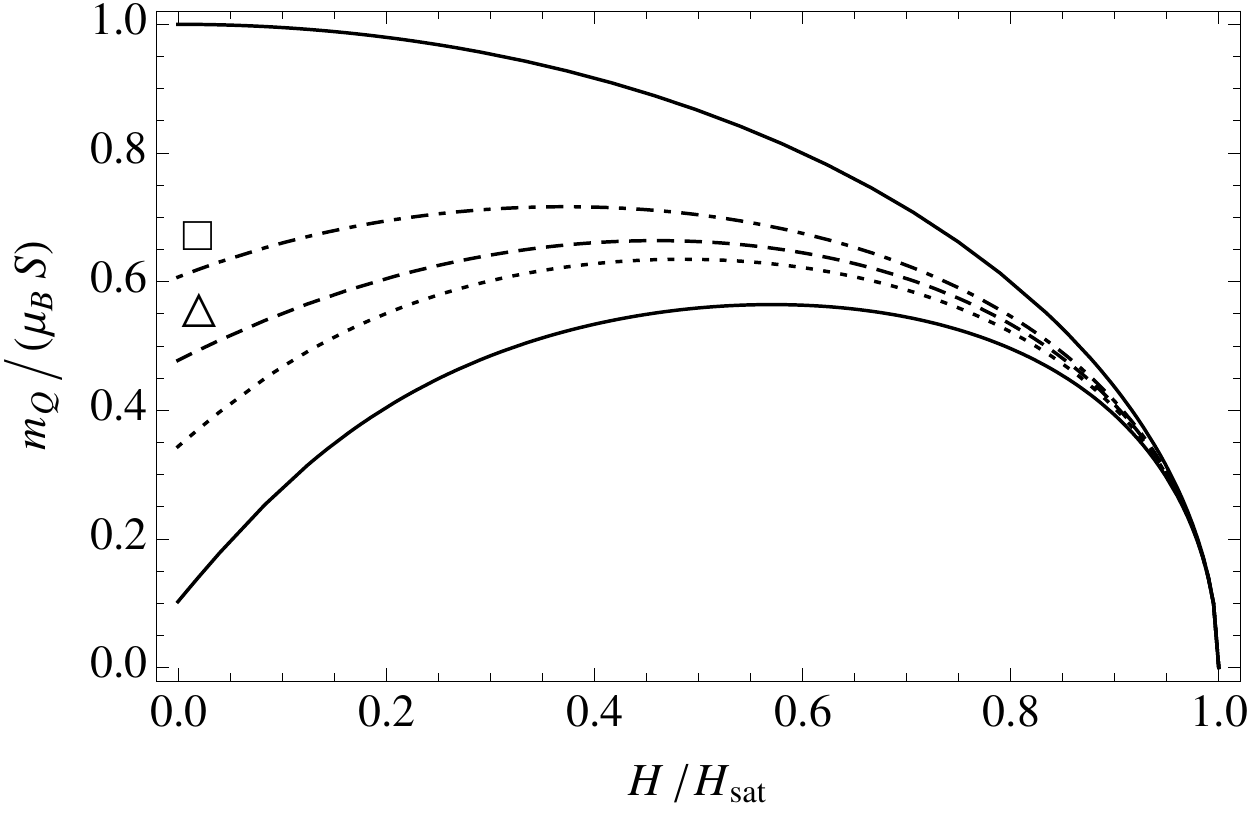}\vspace{0.3cm}
    \includegraphics[width=.9\columnwidth]{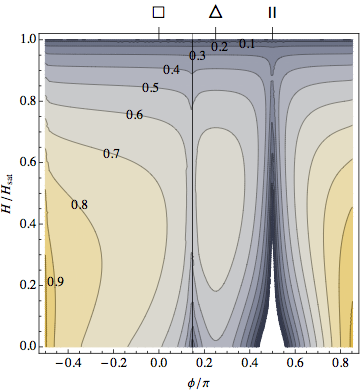}

    \caption{Top: Field dependence of the ordered moment $m_{Q}(H)$
    for $S=1/2$ at different anisotropy angles $\phi$.  Lowest solid
    line: $\phi=0.4\pi$ -- near the crossover to
    quasi-onedimensionality, dashed line: $\phi=\pi/4$ -- isotropic
    triangular antiferromagnet, dotted line: $\phi=0.18\pi$ -- near
    the boundary between spiral and antiferromagnetic phase,
    dot-dashed line: $\phi=0$ -- square-lattice antiferromagnet.  The
    upper (solid) line describes the classical $S\to\infty$ behavior.
    Bottom: $\phi-H$ contour plot of the ordered moment.  Instability
    regions around $\phi/\pi=\tan^{-1}(1/2)/\pi\approx0.15$ and~$0.5$
    become narrower for increasing H. Labels denote $0\leq
    m_Q/(\mu_BS)\leq 1$ in steps of 0.1.}
    \label{fig:ms}
\end{figure}
Turning on a magnetic field, the moments on each site cant towards the
field direction defining our global $z$ axis.  The ordered moment
which is the projection of the former onto the plane perpendicular to 
$z$ can then be written as~\cite{siahatgar:11}
\begin{widetext}
    \begin{equation}
        m_{Q}(H)
        =
        \mu_{\text B}S\sqrt{1-\left(\frac
H{H_{\text{sat}}}\right)^{2}}
        \left\{
        1-\frac1{1-(H/H_{\text{sat}})^{2}}\frac1{2S}
        \int\frac{{\rm d}^{2}q}{V_{\text{BZ}}}\left[
        \frac{A(\vec q)}{\omega(\vec q)}-1
        +\left(\frac H{H_{\text{sat}}}\right)^{2}
        \frac{B(\vec q)}{\omega(\vec q)}
        \left(
        \frac{A(\vec q)-B(\vec q)}{A(0)}-1
        \right)
         \right]\right\}
    \end{equation}
\end{widetext}
up to ${\cal O}(1/S)$, including corrections to the classical canting
angle $\Theta_{\text c}$.  Figure~\ref{fig:ms} displays the field
dependence of the ordered moment for $S=1/2$ and different anisotropy
angles $\phi$.  For comparison also the classical $S\to\infty$ field
dependence is shown (topmost curve).  In Fig.~\ref{fig:mo} for all
four values of $\phi$ the ordered moment at $H=0$ is reduced from its
classical value by quantum fluctuations.  We have chosen
characteristic values for the anisotropy: $\phi=0.4\pi$ (lowest solid
line) is near the crossover to the quasi-onedimensional behavior.  The
dashed curve has $\phi=\pi/4$, which is the isotropic triangular
antiferromagnet starting at $m_Q(H=0)/(\mu_BS)=0.478$.  The dotted
curve with $\phi/\pi=0.18$ is at the AF/spiral boundary.  The
zero-field moments for $\phi/\pi=0.4, 0.18$ are both lower than for
the isotropic trigonal case $\phi=0.25\pi$.  This means that in the
corresponding spiral phase region $m_Q(H=0,\phi)/(\mu_BS)$ is
non-monotonic as function of control parameter $\phi$.  Finally the
dash-dotted curve for $\phi=0$ corresponds to the unfrustrated
square-lattice antiferromagnet.  The moment reduction is strongest
near the quasi-onedimensional region.  Upon increasing $H$, two
effects determine $m_{Q}(H)$: (a) Due to the suppression of quantum
fluctuations by the magnetic field, the total moment is stabilized,
leading to an initial increase of $m_{Q}(H)$.  (b) The total moment is
tilted towards the direction of the magnetic field, leading to a
decrease of $m_{Q}(H)$, which at sufficiently large fields dominates
the field dependence of $m_{Q}(H)$.  In the classical limit,
$m_{Q}(H)$ is determined only by the tilting angle $\Theta_{\text c}$
relative to the field direction.  The initial slope $\lim_{H\to0}
m_{Q}/H$ therefore can be used as a means to determine the anisotropy
angle $\phi$ from neutron diffraction experiments in a magnetic field
as has already been demonstrated for the square-lattice $J_1-J_2$
model \cite{siahatgar:11}.

\section{Magnetization and susceptibility}
\label{sec:magnetic}
\begin{figure}
    \centering
    \includegraphics[width=.9\columnwidth]{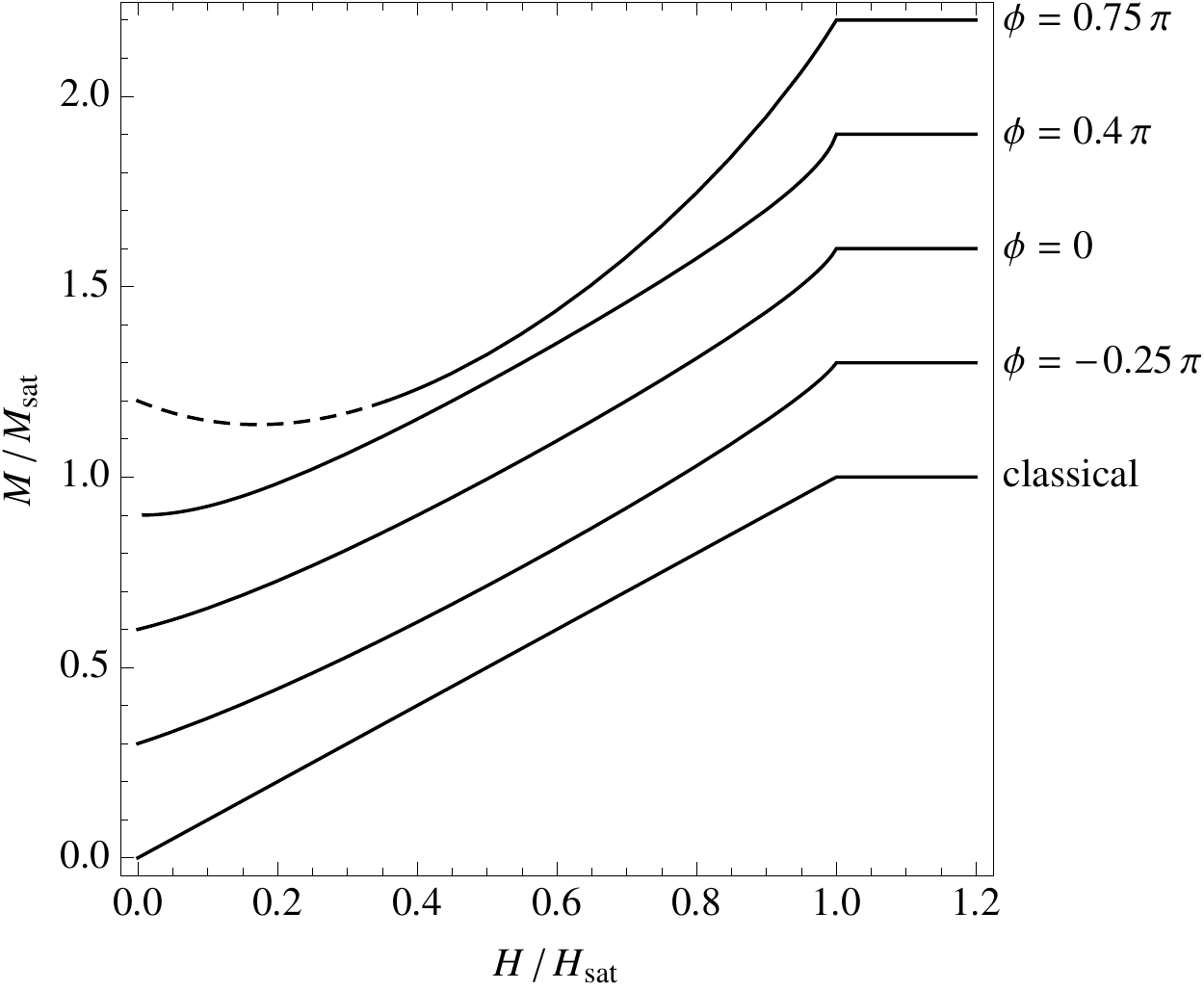}
    \caption{Homogeneous magnetization $M$ for $S=1/2$ in units of
    the saturation magnetization $M_{\text{sat}}=(N/V)g\mu_{\text
    B}S$.  Shown are curves for the classical limit and for different
    values of the anisotropy angle with an offset of $0.3$.  The
    dashed line for $\phi=0.75\pi$ denotes the range where
    fluctuations overcompensate the classical value.  The magnetic
    field is normalized to the (anisotropy-dependent) saturation
    value $H_{\text{sat}}=2SA(0)/(\mu_{0}g\mu_{\text B})$
(Fig.~\ref{fig:ecl}b).}
    \label{fig:magnetization}
\end{figure}
The magnetization $M$ can be derived from the ground-state expectation
value of the $z$ component of the total spin $\vec S=\sum_{i}\vec
S_{i}$.  It is given by
\begin{equation}
    M
    =
    \frac NV
    \frac{(g\mu_{\text B})^2\mu_{0}H}{2A(0)}
    \left[
    1+\frac{1}{2S}\int\frac{{\rm d}^{2}q}{V_{\text{BZ}}}
    \frac{B(\vec q)\left(A(\vec q)-B(\vec q)\right)}{A(0)
    \omega(\vec q)}
    \right]
    \label{eqn:magnetization}
\end{equation}
where $V$ denotes the sample volume.  We note that $\omega(\vec q)$
also depends on $H$, see Eq.~(\ref{eqn:omega}).  For $S=1/2$,
Fig.~\ref{fig:magnetization} displays a plot of $M(H)$ for different
values of the anisotropy angle $\phi$, together with the classical
result.  Both field $H$ and magnetization $M$ are normalized to their
saturation values $H_{\text{sat}}$ and
$M_{\text{sat}}=(N/V)g\mu_{\text B}S$, respectively.  For better
distinguishability, the individual curves are offset by $0.3$.  Deep
within the antiferromagnetic phase (example: $\phi=-\pi/4$), the
corrections to the classical field dependence are small, and increase
with increasing $\phi$.  At $\phi=0$, the field dependence corresponds
to the square-lattice antiferromagnet.  At $\phi=0.4\pi$ or
$J_{2}/J_{1}\approx 3.1$, the border of the stability range of the
ordered moment is reached, and corrections to $M(H)$ lead to a strong
curvature in particular at small fields.  For even larger values of
$\phi$, the fluctuations overcompensate the classical value for $M$,
leading to an unphysical negative magnetization at small fields.  As
an example, we show $M(H)$ for $\phi=3\pi/4$, where the dashed line
indicates this overcompensation. We also note that the magnetization
plateau at $M/M_{\text{sat}}=1/3$ \cite{honecker:99} cannot be
obtained
within LSW since it is due to bound state formation of spin waves.
\begin{figure}
    \centering
    \includegraphics[width=.9\columnwidth]{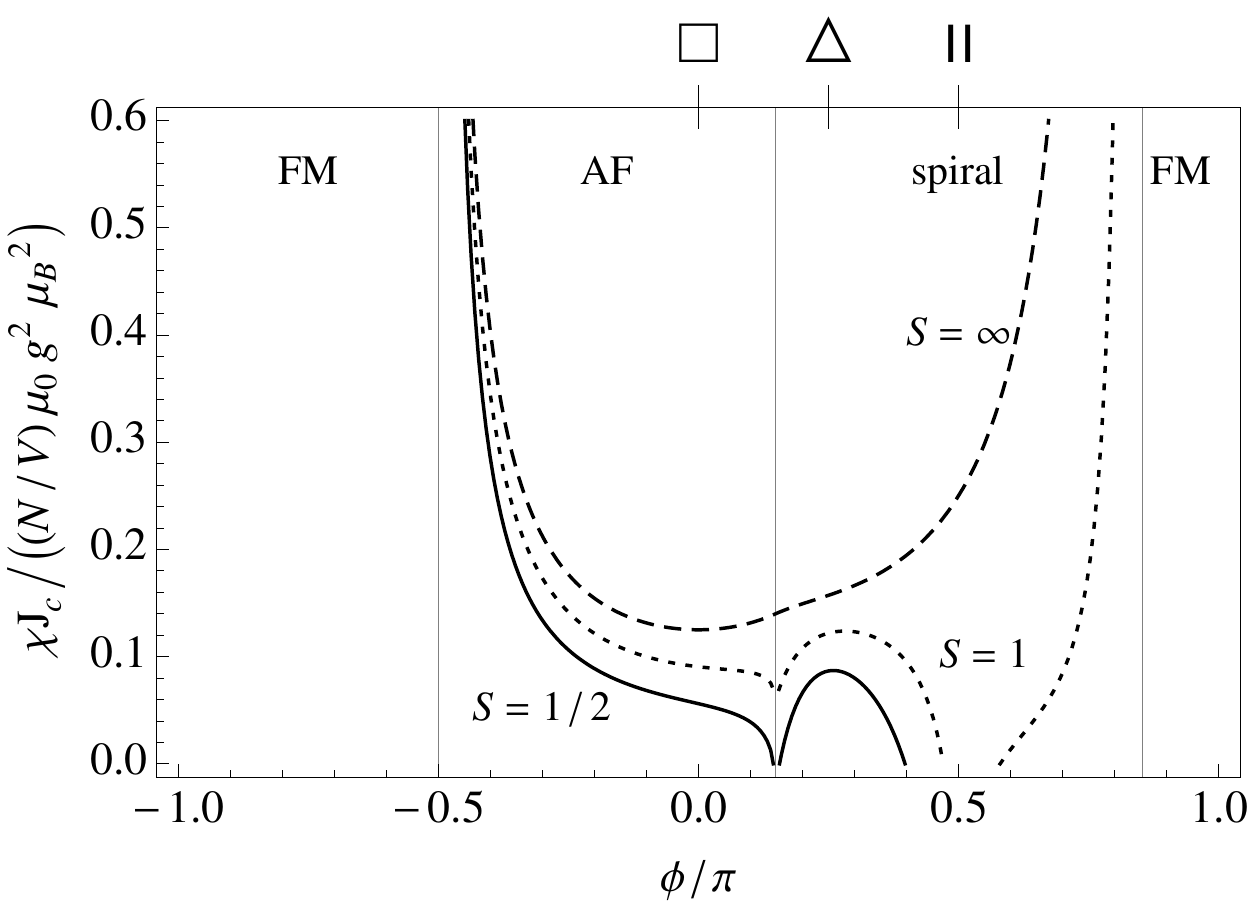}
    \caption{Magnetic susceptibility $\chi=\lim_{H\to0}M/H$ as a
    function of the anisotropy angle $\phi$.  Solid line: $S=1/2$,
    dotted line: $S=1$, dashed line: classical limit $S\to\infty$.}
    \label{fig:chi}
\end{figure}
To illustrate the small-field corrections to the classical case more
clearly, Fig.~\ref{fig:chi} shows a plot of the magnetic
susceptibility $\chi=\lim_{H\to0}M/H$ as a function of the anisotropy
angle $\phi$.  The solid line denotes the anisotropy dependence for
$S=1/2$, the dotted line shows $S=1$, and the dashed line shows the
field dependence for $S\to\infty$.  Already for $S=1$, the unphysical
range of $\phi$ is restricted to values around the one-dimensional
case $\phi=\pi/2$.
\begin{figure*}
    \centering
    \includegraphics[width=\textwidth]{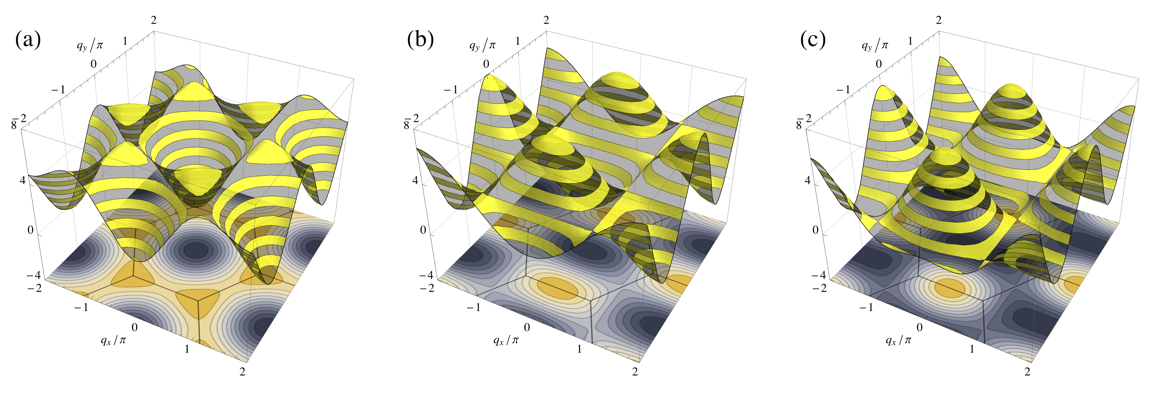}
    \caption{Ferromagnetic spin wave dispersion
$\omega_{\text{sw}}(\vec
    q)$ in units of $J_{\text c}$.  From left to right: (a)
    $J_{1}=J_{2}$ -- isotropic triangular ferromagnet, (b) $J_{2}=0$
    -- square-lattice ferromagnet, (c) $J_{1}\approx-2J_{2}$,
    $J_{2}>0$ -- border with spiral phase.}
    \label{fig:swfm}
\end{figure*}

\section{Excitation spectra}
\label{sec:excitation}

Here we discuss the systematic variation of the LSW excitation
spectrum
when the control parameter $\phi$ is tuned through the phase diagram
of 
Fig.~\ref{fig:pd}. We also emphasize the variation of anisotropic
spin wave
velocities, their measurement may be useful for determination of
$\phi$.

\label{subsec:ferro}

{\em Ferromagnet.} The fully polarized state remains the unchanged
ground state of the Hamiltonian for $J_{1}\le0$ and
$J_{2}/|J_{1}|\le1/2$, however its excitation spectrum strongly
depends on the anisotropy parameter $\phi$.  It is given by
\begin{eqnarray}
    \omega_{\text{sw}}(\vec q)
    &=&
    -4\left(J_{1}+\frac{1}{2}J_{2}\right)
    \\&&
    +4\left(J_{1}\cos\frac{q_{x}}{2}\cos\frac{\sqrt{3}}{2}q_{y}
    +\frac{1}{2}J_{2}\cos q_{x}\right).
    \nonumber
\end{eqnarray}
The dispersion for selected values of the anisotropy parameter $\phi$
is shown in Fig.~\ref{fig:swfm}.  Crossing the border from the Néel
antiferromagnet, $J_{1}$ changes sign, and the dispersion is identical
to that of a one-dimensional ferromagnetic chain (quadratic dispersion
along $q_{x}$ at $\vec q=0$).  Deep within the ferromagnetic region,
the
long-wavelength dispersion around the ordering vector $\vec Q=0$ can
be written as
\begin{equation}
    \omega_{\text{sw}}(\vec q)
    \approx
    \frac{D_{x}}Sq_{x}^{2}+\frac{D_{y}}Sq_{y}^{2},
    \label{eqn:d}
\end{equation}
introducing two stiffness constants $D_{\alpha}$ given by
\begin{equation}
    D_{x}
    =
    -\frac S2\left(J_{1}+2J_{2}\right),
    \quad
    D_{y}
    =
    -\frac32SJ_{1},
\end{equation}
which are positive and identical in the isotropic case $J_{1}=J_{2}$.
The full spectrum for this case is shown in
Fig.~\ref{fig:swfm}a, having a sixfold symmetry around the zeroes at
the Bragg points $\vec Q^{*}=2\pi(n,m/\sqrt3)$ with $n,m\in\mathbb Z$.
Figure~\ref{fig:swfm}b shows the case $J_{2}=0$, which is a
pure square-lattice ferromagnet with exchange constant $J_{1}$.  Apart
from a scaling factor $1/\sqrt3$ for $q_{y}$, this is reflected in the
fourfold-symmetric shape of the spectrum around the Bragg points.  At
the border to the spiral phase for $J_{2}=|J_{1}|/2$ shown in plot
(c), $D_{x}$ vanishes, indicating the instability towards
incommensurate order.

\label{subsec:antiferro}
\begin{figure*}
    \centering
    \includegraphics[width=\textwidth]{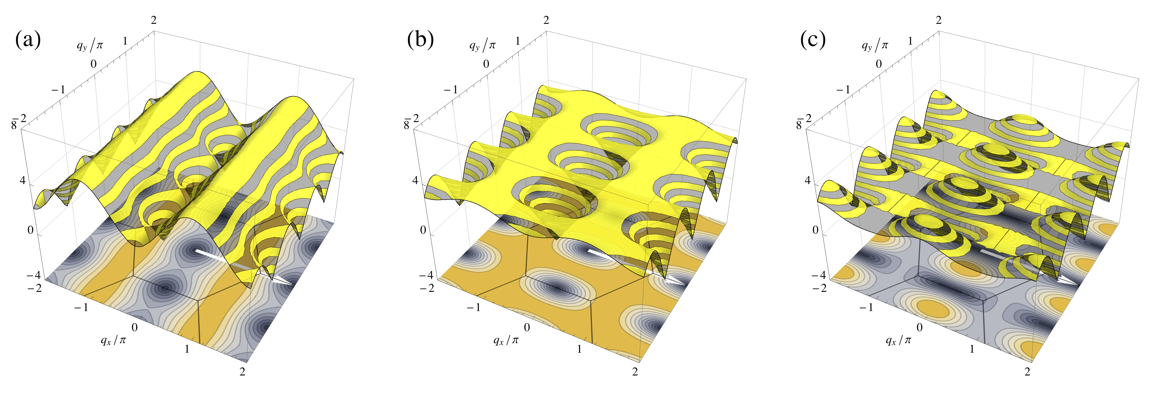}
    \caption{Antiferromagnetic spin wave dispersion
    $\omega_{\text{sw}}(\vec q)$ in units of $J_{\text c}$.  From
left to
    right: (a) $J_{1}=|J_{2}|$ -- deep inside antiferromagnetic
    phase, (b) $J_{2}=0$ -- square-lattice antiferromagnet, (c)
    $J_{1}\approx2J_{2}>0$ -- border with spiral phase.  The white
    arrow denotes the ordering vector $\vec Q=(2\pi,0)$.}
    \label{fig:swaf}
\end{figure*}
{\em Antiferromagnet.} Fig.~\ref{fig:swaf} displays a series of plots
for the spin wave dispersion in the two-sublattice antiferromagnetic
phase.  It is given by
\begin{eqnarray}
    \omega_{\text{sw}}(\vec q)
    &=&
    \sqrt{A^{2}(\vec q)-B^{2}(\vec q)},
    \label{eqn:esw}
    \\
    A(\vec q)
    &=&
    4\left(J_{1}-\frac12J_{2}\right)+2J_{2}\cos q_{x},
    \\
    B(\vec q)
    &=&
    4J_{1}\cos\frac{q_{x}}2\cos\frac{\sqrt3}2q_{y}.
\end{eqnarray}
\begin{figure}
    \centering
    \includegraphics[width=.9\columnwidth]{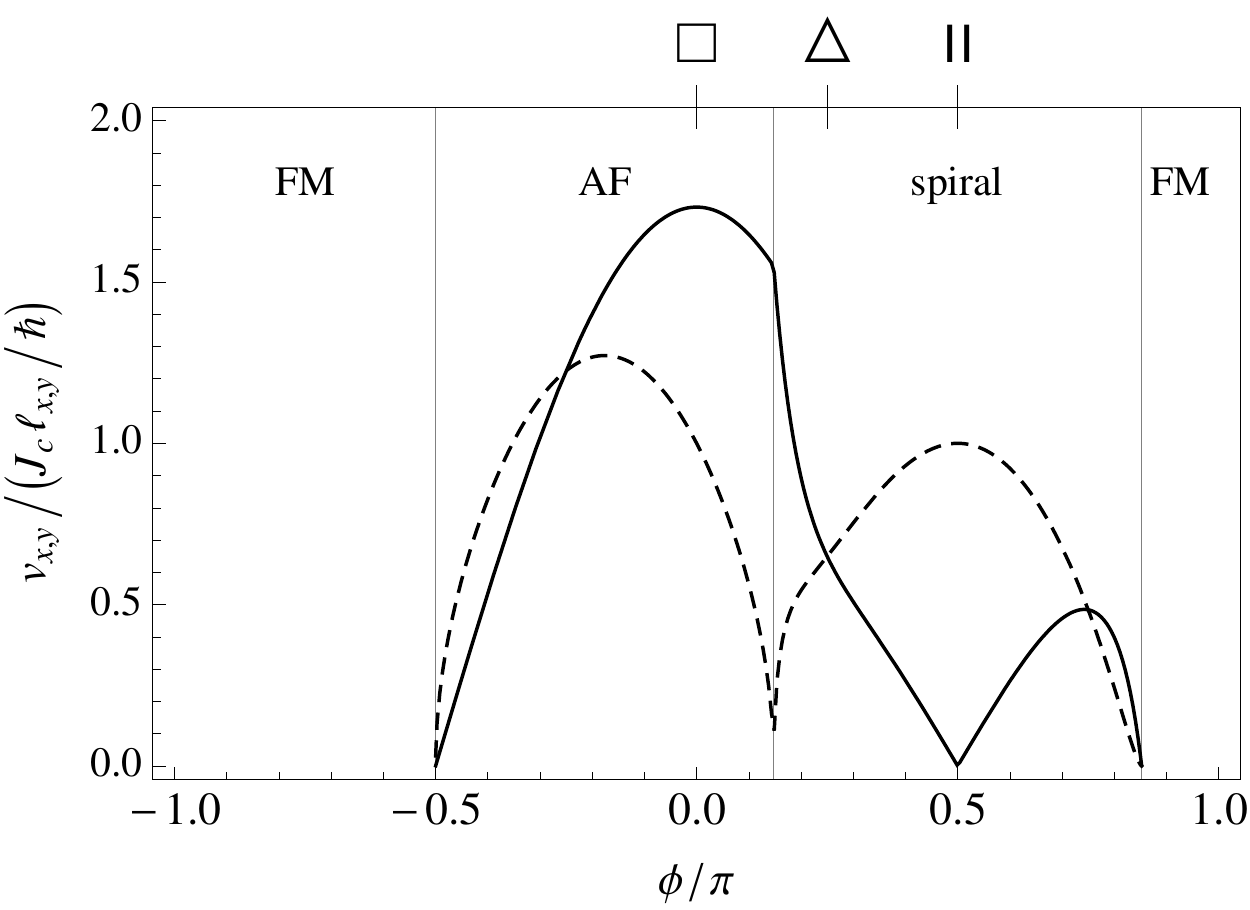}
    \caption{Evolution of the spin-wave velocities $v_{x}$, $v_{y}$ in
    the antiferromagnetic and spiral phases as a function of the
    anisotropy angle $\phi$.  Dashed line: $v_{x}$, solid line:
    $v_{y}$.}
    \label{fig:v}
\end{figure}
As expected, the intra-sublattice dispersion $A(\vec q)$ only depends
on $q_{x}$.  An expansion around the ordering vector $\vec Q$ yields
an expression
\begin{equation}
    \omega_{\text{sw}}^{2}(\vec q)
    \approx
    \left(\frac{\hbar v_{x}}{S\ell_{x}}\right)^2
    \left(q_{x}-Q_{x}\right)^{2}
    +\left(\frac{\hbar v_{y}}{S\ell_{y}}\right)^2
    \left(q_{y}-Q_{y}\right)^{2}
    \label{eqn:v}
\end{equation}
where $\ell_{x,y}$ denote the lattice spacings along $x$ and $y$
directions.  ($\ell_{x}=a$, $\ell_{y}=(\sqrt3/2)a$ for the
geometrically isotropic triangular lattice with lattice constant $a$.)
The anisotropic spin-wave velocities are given by
\begin{equation}
    v_{x}
    =
    \frac{S\ell_{x}}{\hbar}
    2J_{1}\sqrt{1-2\frac{J_{2}}{J_{1}}},
    \quad
    v_{y}
    =
    \frac{S\ell_{y}}{\hbar}
    2\sqrt3J_{1}.
\end{equation}
Fig.~\ref{fig:v} shows the evolution of $v_{x,y}$ as a function of the
anisotropy angle $\phi$.  At the border to the FM phase, $v_{x,y}$
increase from $v_{x}=v_{y}=0$ and the change from
quadratic $\vec q$ dependence from Eq.~(\ref{eqn:d}) to the low-energy
linear dispersion given by Eq.~(\ref{eqn:v}) around the points $\vec
Q^{*}$ is observed.  Towards the inside of the AF phase, these
dispersion cones
stabilize and, for small excitation energies, eventually get nearly
isotropic for $J_{2}=-J_{1}$ or $\phi=-\pi/4$ (Fig.~\ref{fig:swaf}a)
with $v_{x}/\ell_{x}=v_{y}/\ell_{y}=2\sqrt3SJ_{1}/\hbar$.
For high energies however, the flat $q_{y}$ dependence of
$\omega_{\text{sw}}(\vec q)$ remains.  For $J_{2}=0$, the model
describes the familiar square-lattice Néel antiferromagnet with
exchange coupling $J_{1}$, see Fig.~\ref{fig:swaf}b.
Towards the spiral phase, new maxima in $\omega_{\text{sw}}(\vec q)$
develop.  The low-energy cones become soft in $q_{x}$ direction
until at $J_{2}=J_{1}/2$, the spin-wave velocity $v_{x}$ vanishes
(Fig.~\ref{fig:v}).  Note that this does {\em not\/} lead to line
zeros in
the spectrum.  Minima in the antiferromagnetic phase occur both in the
center of the chemical Brillouin zone and at the border of it with
$q_{y}=n\cdot2\pi/\sqrt3$, $n\in\mathbb Z$.

\label{subsec:spiral}
\begin{figure*}
    \centering
    \includegraphics[width=\textwidth]{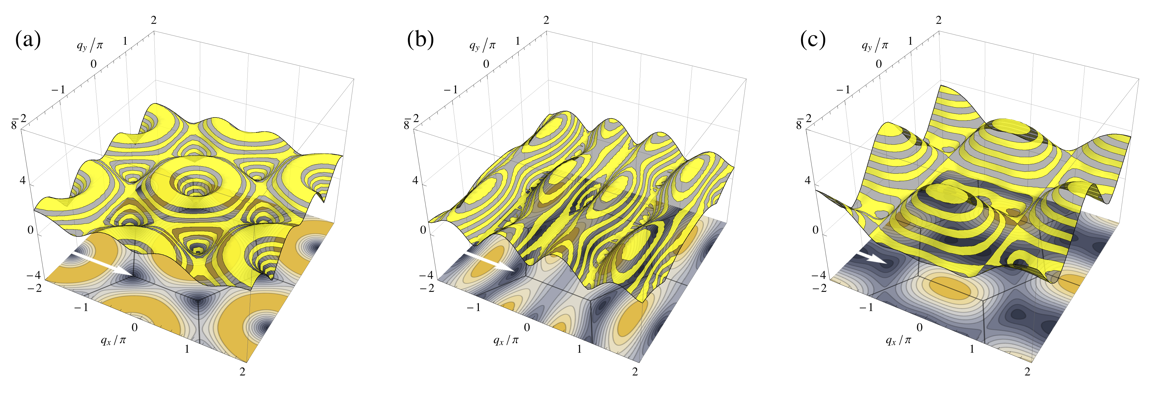}
    \caption{Spin wave dispersion $\omega_{\text{sw}}(\vec q)$ in
units of $J_{\text c}$ in
    the spiral phase.  From left to right: (a) $J_{2}=J_{1}$ --
    isotropic triangular antiferromagnet, (b) $J_{2}/J_{1}\approx3$ --
    border to disordered region, (c) $J_{2}=-J_{1}>0$ -- spiral with
    ferromagnetic $J_{1}$.  For better visibility, the white arrows
    showing the ordering vector $\vec Q$ are shifted by $\Delta\vec
    Q=-(2\pi,2\pi/\sqrt3)$. The case (b) corresponds approximately to
\CCC{} (see Table~\ref{tbl:compounds}).}
    \label{fig:swsp}
\end{figure*}
{\em Spiral phase.} Between $\phi=\tan^{-1}(1/2)$ and
$\phi=\pi-\tan^{-1}(1/2)$, the ordering vector in general is
incommensurate and has the form $\vec Q=(Q_{x},0)$ with $0\le
Q_{x}\le2\pi$.  The spin-wave spectra obtained are plotted in
Fig.~\ref{fig:swsp}.  The dispersion is again given by
Eq.~(\ref{eqn:esw}), and the coefficients $A(\vec q)$ and $B(\vec q)$
retain the general forms given by Eqs.~(\ref{eqn:a})
and~(\ref{eqn:b}).  Expanding $\omega_{\text{sw}}(\vec q)$ around the
ordering vector leads to a linear $\vec q$ dependence around the
dispersion minima given by Eq.~(\ref{eqn:v}).  The spin-wave
velocities in this case are
\begin{eqnarray}
    v_{x}
    &=&
    \frac{S\ell_{x}}{\hbar}2J_{2}
    \left(1+\frac{J_{1}}{2J_{2}}\right)^{3/2}
    \nonumber\\
    &&{}\times
    \sqrt{1-\frac32\frac{J_{1}}{J_{2}}
    +\left(\frac{J_{1}}{J_{2}}\right)^{2}
    -\frac14\left(\frac{J_{1}}{J_{2}}\right)^{3}},
    \\
    v_{y}
    &=&
    \frac{S\ell_{y}}{\hbar}\sqrt 3J_{1}
    \left(1+\frac{J_{1}}{2J_{2}}\right)
    \sqrt{1-\frac{J_{1}}{J_{2}}
    +\frac12\left(\frac{J_{1}}{J_{2}}\right)^{2}}.
\end{eqnarray}
In the isotropic case ($J_{1}=J_{2}=J_{\text c}/\sqrt{2}$), we have
$\hbar v_{x}/\ell_{x}=\hbar v_{y}/\ell_{y}=(3\sqrt{3}/4)SJ_{\text
c}\approx1.3\,SJ_{\text c}$ (Fig.~\ref{fig:v}).  Increasing $\phi$
from the border to the antiferromagnetic phase, we have a crossover
from the fourfold symmetry of the Néel antiferromagnet to the sixfold
symmetry of the isotropic triangular magnet, see Fig.~\ref{fig:swsp}a.
Low-energy cones with linear $\vec q$ dependence develop in the center
and at the corners of the Brillouin zone.  Upon further increase of
$\phi$, the sixfold symmetry is transformed into the twofold-symmetric
one-dimensional pattern, see Fig.~\ref{fig:swsp}b.  For ferromagnetic
$J_{1}<0$ (but still antiferromagnetic $J_{2}$ and an in general
incommensurate ordering vector), the excitation spectrum becomes
similar to what we have found in the ferromagnetic phase.

\section{Discussion}
\label{sec:discussion}

\begin{table*}
    \caption{Exchange parameters of some compounds with frustrated
    triangular spin structure}
    \vspace{0.3cm}
    \begin{ruledtabular}
        \begin{tabular}{ccccccccc}
            compound
            &
            $S$
            &
            $J_{2}/J_{1}$
            &
            $\phi/\pi$
            &
            $J_{\text c}/k_{\text B}$
            &
            $Q_{\text{cl}}/\pi$
            &
            $m_{Q}^{\text{sw}}/(\mu_{\text B}S)$ (exp)
            &
            $\left[\left.(v_{x}/\ell_{x})\right/
            (v_{y}/\ell_{y})\right]_{\text{sw}}$
            &
            reference
            \\
            \hline
            Cs$_2$CuCl$_{4}$
            &
            $1/2$
            &
            $3.31$
            &
            $0.41$
            &
            $4.9\,\text K$
            &
            $1.10$
            &
            $0.043$
            &
            3.93
            &
            \cite{zvyagin:13}
            \\
            Cs$_2$CuBr$_{4}$
            &
            $1/2$
            &
            $2.44$
            &
            $0.38$
            &
            $16.1\,\text K$
            &
            $1.13$
            &
            $0.20$
            &
            2.86
            &
            \cite{zvyagin:13}
            \\
            NiGa$_{2}$S$_{4}$
            &
            $1$
            &
            $1$\footnote{antiferromagnetic next-nearest neighbor
            interaction $J_{3}>|J_{1}|,|J_{2}|$ dominates\cite{nakatsuji:10}}
            &
            $-0.75$
            &
            $6.45\,\text K$
            &
            $-$
            &
            $-$
            &
            $-$
            &
            \cite{nakatsuji:10}
            \\
            PdCrO$_{2}$
            &
            $3/2$
            &
            $1$
            &
            $0.25$
            &
            $42.7\,\text K$\footnote{from $v_{x}=v_{y}=3200\,\text
            m/\text s$ \cite{takatsu:09}}
            &
            $4/3$
            &
            $0.826$ ($0.666$)
            &
            $1$
            &
            \cite{takatsu:09,mekata:95}
        \end{tabular}
    \end{ruledtabular}
      \protect\label{tbl:compounds}
\end{table*}
We begin the further discussion of our results with the classical
phase diagram.  It results from the classical ground state energies
plotted in Fig.~\ref{fig:ecl}.  The phase boundaries appear at the
same values of $\phi$ as in the case of the square lattice
$J_{1}-J_{2}$ model.  However in the range $0.15\leq\phi/\pi\leq0.85$
the ground state for the triangular model is a generally
incommensurate (IC) spiral whereas the square lattice model has a
commensurate (C) columnar AF ground state.  The reason may be
understood from Fig.~\ref{fig:ecl}: In the above range the lowest
(dotted) energy curve of the spiral phase is only slightly lower than
the one for the columnar AF (short-dashed) which shows a simple
$-J_c\sin\phi=-J_2$ dependence.  They touch at $\phi/\pi=0.5$ for the
decoupled chain model.  In the case of the square lattice model,
however, there is one diagonal $J_2$ exchange bond more, therefore the
(short dashed) $E_{cl}(\phi)$ curve for the columnar AF doubles its
amplitude to $-2J_c\sin\phi=-2J_2$ and now lies below the spiral phase
in the above range of $\phi$, thus leading to a columnar AF ground
state for the square lattice model.  This explains the main difference
of the square lattice and triangular classical phase diagrams.  The
latter is shown in Fig.~\ref{fig:pd}.

It is naturally important to investigate how many features of the
classical phase diagram are modified or changed radically by the
effect of quantum fluctuations.  This relates only to the AF and
spiral phases since in the FM phase quantum fluctuations are absent.
We have included their effect in order (1/S) within conventional LSW
approximation.  It must be clearly stressed that the latter breaks
down when there is no long range magnetic order.  This happens in
possibly three regions of the phase diagram: At the boundaries of the
spiral region to the AF and FM phase and in the center of it ($J_1=0$)
where decoupled chains lead lead to a quasi-1D spin liquid state to be
discussed further below.

First we note that, as in the square lattice
model the phase boundaries are unchanged in LSW approach when zero
point fluctuations are included
in the ground state energy. However, in the triangular model we have
the additional aspect that the spiral magnetic structure itself,
i.e., the spiral propagation vector {\bf Q} is 
modified by the zero point contributions to the ground state energy
as derived in Sec.~\ref{sec:fluctuations}(Fig.~\ref{fig:q}).

More pronounced quantum effects occur in the the ground state energy
which is shown in Fig.~\ref{fig:egs}a.  The quantum corrected $E_{gs}$
curve (full line) lies considerably below the classical on one (dashed
line) in the AF and spiral region (as already mentioned there are no
corrections in the FM state).  In particular we note that even for the
1D case ($\phi/\pi=0.5$ or $J_{1}=0$) where nominally spin wave theory
is not valid we obtain an excellent value as compared to the exact
Bethe ansatz result \cite{bethe:31} energy which is only slightly
lower.  This may be justified by the fact that although no strict long
range order exists quasi-long range order signified by an algebraic
decay of the correlation function is present in quasi-1D AF spin
chains.  Therefore the long-wavelength spin excitations which are most
important for the zero point corrections are qualitatively reasonably
described by spin waves.

Now we summarize the most important quantum effects concerning the
renormalization of the ordered moment $m_Q$ that have been derived in
Sec.~\ref{sec:moment} (Fig.~\ref{fig:mo}a).  For comparison we also
present the previous results for the $J_1-J_2$ square lattice model in
Fig.~\ref{fig:mo}b.  In Fig.~\ref{fig:mo}a in the AF regime the
variation is smooth starting at the classical value $m_Q/(\mu_BS)=0.5$
at the FM boundary and passing through the unfrustrated (square
lattice) value of $m_Q/(\mu_BS)=0.3$ at $\phi=0$ ($J_{2}=0$) and
vanishing logarithmically at the spiral phase boundary point
($\phi/\pi\approx0.15$ or $J_{2}=0.5J_{1}$).  For larger $\phi$ in the
spiral region the moment recovers and reaches a maximum value of
$m_Q/(\mu_BS)=0.24$ in the frustrated isotropic triangular model
($\phi/\pi=0.25$ or $J_{1}=J_{2}$).  This is in reasonable agreement
with more exact DMRG results \cite{white:07} giving
$m_Q/(\mu_BS)\simeq 0.20$.  Then $m_Q$ decreases rapidly with
increasing $\phi$ and is destroyed in a finite region around the 1D
chain model ($\phi/\pi=0.5$ or $J_{1}$=0) signifying the onset of a
quasi-1D spin liquid state.  The width of the sector where $m_Q$
vanishes in LSW approximation is given by the shaded area in
Fig.\ref{fig:pd}.  For still larger $\phi$ the ordered moment $m_Q$ is
stabilized again and reaches the classical value at the FM boundary.

The effect of quantum fluctuations on the ordered moment can be tuned
by applying a magnetic field.  A non-monotonic behavior as function of
$H$ was found in Fig.~\ref{fig:ms} due to reduction of zero point
fluctuations by the polarization, quite similar to the $J_1-J_2$ model
\cite{siahatgar:11}.  In contrast, however, the triangular model also
exhibits a non-monotonic behavior of the zero-field moment
$m_Q(H=0,\phi)$ as function of control parameter $\phi$ around the
isotropic point ($\triangle$) with $\phi/\pi=0.25$ as seen in
Fig.~\ref{fig:ms}.

The homogeneous magnetization and the susceptibility for the
triangular model were found to be very anomalous in the spiral phase
(Figs.~\ref{fig:magnetization},\ref{fig:chi}).  LSW approximation
breaks down for these quantities in more than half of the spiral
sector $0.4\leq \phi/\pi \leq 0.85$.  It is remarkable that the
homogeneous magnetization M (and $\chi$) is not recovered close to the
FM sector even though the the ordered moment is stabilized again
beyond the 1D point ($\parallel$) for $\phi/\pi >0.6$
(Fig.~\ref{fig:mo}a).

Thus in LSW treatment there are three instability regions of magnetic
order in the anisotropic triangular model of Fig.~\ref{fig:mo}a.  1)
At the AF-spiral phase boundary point.  In LSW approach we cannot
identify a sizable finite sector size of the instability.  This is in
contrast to the square lattice $J_{1}-J_{2}$ model
(Fig.~\ref{fig:mo}b) where the instability for $J_{1}=J_{2}$ occurs in
a finite sector \cite{schmidt:10}.  The true ground state in that
sector is likely of the nonmagnetic staggered dimer type
\cite{singh:99} and this has also been suggested for the triangular
model from finite temperature series expansions \cite{weihong:99}.
However in this case a finite sector for the nonmagnetic state shifted
to slightly larger $\phi/\pi\simeq 0.19$ is predicted.  This is
supported by modified spin wave theory and exact diagonalization
results \cite{hauke:13}.  2) The quasi -1D spin fluctuations around
$\phi/\pi\simeq 0.5$ lead to an extended region of a quasi-1D spin
liquid state with algebraic spin correlations.  This is opposite to
the situation in the square lattice $J_{1}-J_{2}$ model which achieves
the unfrustrated 2D HAF value $m_Q/(\mu_B)=0.3$ at this point due to
decoupled unfrustrated (J$_2$ only) sublattices.  3) at the spiral-FM
phase boundary the staggered moment $m_Q$ is stable, however
homogeneous magnetization and susceptibility behave very anomalous
indicating again a tendency to instability.  This behavior of $M$ and
$\chi$ is similar to the square lattice $J_{1}-J_{2}$ model, although
there the instability range in $\phi$ is smaller.  On the other hand
for the square lattice model the ordered moment itself becomes
unstable at the equivalent columnar AF-FM boundary
(Fig.~\ref{fig:mo}b).  In this case it is known that a non-magnetic
spin-nematic phase \cite{shannon:06} is established in this regime.

As the main purpose of this work is to give a survey of ground states
and excitations for all triangular anisotropies we now further
discuss the variation of the excitation spectrum in LSW theory for
the full range of $\phi$. In each of the magnetically ordered sectors
we choose three typical values of $\phi$ for the 2D spin wave
dispersion and present it both as 3D topographic plot and in shaded
contours in the $q_x-q_y$ plane.

In Fig.~\ref{fig:swfm} we present it for the ferromagnetic regime.
In the underlying contour plot the dark regions represent locations
in the BZ where the spin wave energy is low and the bright (yellow)
regions correspond to high energies. Apparently when $\phi$
approaches the spiral phase boundary the BZ regions with low energy
excitations become large.
This  continues into the adjacent spiral phase and may play a role in
the extended instability of the low-field magnetization or
susceptibility discussed above.

The AF spin wave dispersions are shown in  In Fig.~\ref{fig:swaf}.
They exhibit a continuous change of anisotropy characteristics as
function of $\phi$. Deep inside the AF phase (a) and for the square
lattice case (b) the low energy spin waves are centered around zone
center and zone boundary points in the BZ whereas at the border to
the spiral phase they are located in large overlapping zone boundary
regions (c).

The results for the spiral phase are presented in Fig.~\ref{fig:swsp}.
The low energy spin waves are now located around incommensurate
ordering wave vectors {\bf Q} in the BZ indicated by white arrows.  Of
particular interest is the case (b) with $J_{2}/J_{1}\approx 3$ or
$\phi/\pi \approx 0.4$ which is close to the anisotropy ratios of
\CCC{} and \CCB{} (Table~\ref{tbl:compounds}).  In this case overlapping
regions of low energy spin waves along $q_y$ direction indicate a
large renormalization of the moment.  Indeed from Fig.~\ref{fig:mo}
one concludes a moment reduction to $m_Q/(\mu_BS)=0.043$ in LSW
approximation.  Thus \CCC{} is at the very limit of applicability of
spin wave theory \cite{veillette:05,veillette:05c}.  A slight
increase of $\phi$ would push it to the nonmagnetic quasi-1D spin
liquid regime.  Of course this discussion is oversimplified as in the
real compound there are further interactions like a
Dzyaloshinskii-Moriya term and inter-plane exchange \cite{coldea:02}.
The latter has the tendency to stabilize the magnetic order.  Beyond
the quasi 1D spin liquid state the ordered moment is reestablished
(Fig.~\ref{fig:mo}).  The spin wave dispersion in this regime (c) has
however still large {\bf q} space regions with small energies.  This
may be the reason why the homogeneous magnetization and susceptibility
is not stabilized again (Fig.~\ref{fig:magnetization}).

Finally we discuss the relation of our survey using LSW theory to
other analytical and numerical results.  This is confined to the
sector of AF/spiral phases studied before whose existence is predicted
by all methods.  The precise values for $\phi$ at which the phase
boundaries occur and the position and size of nonmagnetic sectors,
however, depends on the method used.

In LSW theory the results are simple because magnetic phase boundaries
are identical to the classical ones (Table~\ref{tbl:ecl}).  In
particular the AF/spiral boundary is at
$\phi=\tan^{-1}(1/2)\approx0.15\pi$ and the ordered moment is
vanishing only at this point (i.e. there is no finite size of a
nonmagnetic sector in Fig.~\ref{fig:mo}a as opposed to the square
lattice model in Fig.~\ref{fig:mo}b at the same value of $\phi$).  The
spiral sector extends up to $\phi/\pi\approx0.42$ where the moment
vanishes due to the appearance of the quasi-1D spin liquid.

In modified spin wave theory (MSW) \cite{hauke:11,hauke:13} the AF
phase is stabilized to larger values of $\phi/\pi\approx0.20$ where,
in contrast to LSW theory the nonmagnetic state becomes stable in a
finite interval up to $\phi/\pi\approx0.23$.  It is followed by the
spiral phase which achieves its maximum staggered moment $m_{\vec
Q}/(\mu_{\text B}S)\approx0.685$ for the isotropic case
$\phi/\pi=0.25$ with the $120^\circ$ structure, as in LSW theory.  The
latter has $m_{\vec Q}/(\mu_{\text B}S)\approx0.478$, in better
agreement with the DMRG value $m_{\vec Q}/(\mu_{\text
B}S)\approx0.41$.  The spiral magnetic phase then is stable only in a
comparatively small interval up to $\phi/\pi\approx0.32$ where already
the quasi-1D spin liquid appears.  Thus for the exchange anisotropy
commonly used for \CCC{} (Table~\ref{tbl:compounds}) MSW theory
predicts this compound would have a nonmagnetic ground state in a
purely 2D model \cite{hauke:13}.

In addition the AF/spiral boundary region has been studied before
using a numerical dimer series expansion technique~\cite{weihong:99}.
Here also the AF phase is stabilized to larger values
$\phi/\pi\approx0.19$ followed by a finite dimer spin liquid phase
interval up to a value $\phi/\pi\approx0.24$ obtained from vanishing
staggered moment or spin gap, respectively.  For larger $\phi$ values
the spiral phase is again stabilized, however the maximum moment
$m_{\vec Q}/(\mu_{\text B}S)\approx0.18$ is achieved only for an
incommensurate spiral phase with $\phi/\pi\approx0.34$, considerably
larger than in LSW or MSW theory where it occurs for the $120^\circ$
structure of the isotropic case with $\phi/\pi=0.25$.  The quasi-1D
spin liquid boundary is difficult to determine with this method, but
the staggered moment vanishes at similar values $\phi/\pi\approx0.44$
comparable to the LSW result.
 
Thus from these investigations beyond LSW theory one concludes the
following general trends in the AF/spiral region: The stability of the
AF phase is pushed to larger $\phi$ and a finite spin liquid interval
appears between $\phi/\pi\approx0.19\ldots0.20$ and $\phi/\pi\approx
0.23\ldots0.24$.  In addition MSW predicts an earlier onset of the
quasi-1D spin liquid regime.  Thus the stability region of the spiral
magnetic phase generally shrinks as compared to the LSW case.  In
particular for the exchange ratio commonly used for \CCC{}
(corresponding to $\phi/\pi=0.41$) LSW theory and dimer series
expansion still lead to a marginally ordered spiral magnetic state
whereas MSW theory predicts already a disordered state for the purely
2D model without interlayer coupling.

\section{Summary and conclusion}
\label{sec:conclusion}

In this work we have given a global survey of the effect of quantum
fluctuations in triangular spin lattices with anisotropic ferro-and
antiferromagnetic exchange.

We have explored the phase diagram of the triangular $J_1$-$J_2$ model
in the full range of the control parameter $\phi=\tan^{-1}(J_{2}/J_{1})$
using the linear spin wave method.  The contribution of zero point
fluctuations to ground state energy and wave vector, ordered moment,
magnetization and susceptibility has been investigated.  In particular
we point out that the ordered moment shows non-monotonic dependence
both on the field and on the control parameter $\phi$ close to the
isotropic triangular system.  We also made a systematic comparison
with the related square lattice $J_1$-$J_2$ model concerning the
possible phases and their stability as well as their ordered moment
reduction by quantum fluctuations.

Furthermore we discussed the characteristic anisotropies of spin wave
dispersion as the control parameter $\phi$ is tuned through the phase
diagram.  The instability of the ordered moment for the AF/spiral
boundary and the quasi-1D `spin liquid' region is found to be
associated with low energy modes in large areas of the BZ. The spin
wave velocities as function of $\phi$ are very anisotropic, in
particular in the spiral phase.  This may be used for a determination
of the exchange ratio.  A further diagnostic tool fur this purpose is
the (low-) field dependence of the ordered moment which was found to
depend strongly on the control parameter $\phi$.  Naturally the LSW
method used here is not suitable to discuss the region of moment
instability or spin-liquid regions.  More sophisticated numerical and
analytical methods as mentioned in the introduction have to be use for
this purpose.

\bibliography{lsw-triangle}

\end{document}